\documentclass[traditabstract]{aa}
\usepackage{amsmath}

\usepackage{epsfig}
\usepackage{psfig}
\usepackage{graphicx}

\usepackage{natbib}
\usepackage{txfonts}

\bibpunct{(}{)}{;}{a}{}{,} %% natbib format for A&A and ApJ

\begin{document}

\title{SED-dependent galactic extinction prescription for Euclid and future cosmological surveys}

\author{
Audrey Galametz\inst{1}, Roberto Saglia\inst{1}, St{\'e}phane Paltani\inst{2}, Nikolaos Apostolakos\inst{2}, Pierre Dubath\inst{2}
}

\institute{
MPE, Max-Planck-Institut f{\"u}r Extraterrestrische Physik, Giessenbachstrasse, D-85741 Garching, Germany [contact: galametz@mpe.mpg.de]
\and{Department of Astronomy, University of Geneva, ch. d'{\'E}cogia 16, CH-1290 Versoix, Switzerland}
}

\abstract
{The outcome of upcoming cosmological surveys will depend on the accurate estimates of photometric redshifts. In the framework of the implementation of the photometric redshift algorithm for the ESA {\it Euclid} Mission, we are exploring new avenues to improve current template-fitting methods. This paper focusses in particular on the prescription of the extinction of a source light by dust in the Milky Way. Since Galactic extinction strongly correlates with wavelength and photometry is commonly obtained through broad-band filters, the amount of absorption depends on the source intrinsic spectral energy distribution (SED), a point however neglected as the source SED is not known a-priori. A consequence of this dependence is that the observed $E_{\rm B-V}$ ($= A_{\rm B} - A_{\rm V}$) will in general be different from the $E_{\rm B-V}$ used to normalise the Galactic absorption law $k_{\rm \lambda}$ ($= A_{\rm \lambda} / E_{\rm B-V}$). Band-pass corrections are thus required to adequately renormalise the law for a given SED. In this work, we assess the band-pass corrections of a range of SEDs and find they vary by up to $20$\%. We have investigated how neglecting these corrections biases the calibration of dust into reddening map and how the scaling of the map depends of the sources used for its calibration. We derive dust-to-reddening scaling factors from the colour excesses of $z < 0.4$ SDSS red galaxies and show that band-pass corrections predict the observed differences. Extinction corrections are then estimated for a range of SEDs and a set of optical to near-infrared filters relevant to {\it Euclid} and upcoming cosmological ground-based surveys. For high extinction line-of-sights ($E_{\rm B-V} > 0.1$, $\sim8$\% of the {\it Euclid} Wide survey), the variations in corrections can be up to $0.1$~mag in the `bluer' optical filters ($ugr$) and up to $0.04$~mag in the near-infrared filters. We find that an inaccurate correction of Galactic extinction critically affects photometric redshift estimates. In particular, for high extinction lines of sights and $z < 0.5$, the bias (i.e.~the mean $\Delta z = z_{\rm phot} - z_{\rm real}$) exceeds $0.2\%(1+z)$, the precision required for weak-lensing analyses. Additional uncertainty on the parametrisation of the Milky Way extinction curve itself further reduces the photometric redshift precision. We propose a new prescription of Galactic absorption for template-fitting algorithms which takes into consideration the dependence of extinction with SED.}


\authorrunning{Galametz, A. et al.}

\titlerunning{SED-dependent Galactic Extinction Prescription for Euclid}

\maketitle

\section{Introduction}

{\it Photometric redshift precision for weak lensing studies}

The new generation of wide sky surveys such as the ground-based Dark Energy Survey (DES; the Dark Energy Survey Collaboration 2016)\nocite{DES2016}, the Kilo Degree Survey (KiDS; de Jong et al.~2013)\nocite{deJong2013} and the Large Synoptic Survey Telescope (LSST; Ivezi\'c et al.~2008)\nocite{Ivezic2008} survey or the upcoming optical-to-near-infrared ESA {\it Euclid} mission will map the extragalactic sky in multiwavelength photometric bands for a large number of galaxies. One of the primary science goals of these surveys is to constrain the dark energy equation of state by means of powerful cosmological probes including weak gravitational lensing by large-scale structures (cosmic shear). The applications of weak lensing analyses for high precision cosmology demand both an accurate measurement of the source distortion and a statistical knowledge of their distance. Given the few $10^9$ galaxies expected to lie within the survey footprints, these studies will heavily rely on high quality photometric redshifts (photo-z hereafter). 

To measure the effect of dark energy with cosmic time, a number of works have adopted the weak lensing tomography approach \citep[e.g.~][]{Hu1999,Hoekstra2002} where the galaxies are binned in tomographic bins. The weak lensing signal is derived from the cross-correlation of the source distortions between redshift planes, thus providing information on the distribution of mass along the line of sight. The weak lensing tomography approach requires photometric redshifts of high precision to avoid tomographic bin overlaps and to allow an accurate determination of the mean redshift of sources in each bin. This leads to performance requirements on both the photometric redshift scatter and bias. The photometric redshift accuracy required to achieve the precisions on the cosmological parameters set by current cosmological surveys is however spectacularly challenging \citep{Bordoloi2010}. Various studies have for example determined that a requirement of the order of a few percent in the dark energy equation of state parameter $w$ would translate into a scatter $\sigma_{\rm z} < 0.05(1+z)$ and a bias $\langle z \rangle < 0.2\%(1+z)$ \citep[][and references therein]{Ma2006, Hearin2010}. 

The performance of photometric redshift techniques is continuously improving. Photometric redshifts were for example derived for sources in the CANDELS \citep{Grogin2011} GOODS-south field from the multi-wavelength $17$-band photometric catalogue of \citet{Guo2013}. \citet{Dahlen2013} combined the outputs of a number of widely used template-fitting codes (using for instance a simple median or a hierarchical Bayesian approach) and reached a scatter in their photo-z estimates of $\sigma_{\rm z} \sim 0.025(1+z)$ and a bias of $\langle z \rangle < 0.7\%(1+z)$. These results are encouraging, but more work is certainly needed to achieve the ambitious goals set by the new generation of cosmological surveys such as the {\it Euclid} mission. The all (or half) sky surveys mentioned earlier will also evidently lack the richness of multiwavelength data available in deep extragalactic surveys such as GOODS-south. 

In this paper, we focus on the minimisation of a source of photometric bias that originates from an often simplistic treatment of Galactic absorption.

\vspace{4mm}

{\noindent \it Galactic extinction maps}

The interstellar dust of the Milky Way absorbs the UV-to-near-infrared light and contaminates observations of extragalactic objects. Galactic reddening maps are therefore used to correct the observed photometric measurements of distant sources. A number of works have derived reddening maps (commonly referred to as $E_{\rm B-V}$ maps) using the colour excesses of stars or standard cosmological sources with well-known intrinsic spectral energy distribution, in other words with low intrinsic colour dispersion such as quasars, luminous red galaxies etc. These studies assume that the dust properties of the Galactic interstellar medium are homogenous across the Milky Way and linearly rescale either HI column density maps (Burstein \& Heiles 1978)\nocite{Burstein1978} or dust column density maps to match the observed extinction of their calibration sources.

One of the most widely used Galactic reddening maps was derived by Schlegel, Finkbeiner \& Davis (1998; SFD98 hereafter)\nocite{Schlegel1998} from a dust thermal emission map in the far-infrared, the $100\mu$m DIRBE/{\it IRAS}-combined, point-source removed, $6.1\arcmin$ resolution map. The dust column density $D$ was linearly rescaled to a reddening estimate $E_{\rm B-V}$ by a constant calibration coefficient $p$ following $E_{\rm B-V} = pD$. The scaling factor $p$ was estimated using the colour excess estimates of $\sim 100$, $z < 0.05$ brightest cluster galaxies (BCG) and $\sim 400$ field elliptical galaxies. SFD98 adopted the Milky Way absorption law functional form of \citet{O'Donnell1994} in the visible and Cardelli, Clayton \& Mathis (1989)\nocite{Cardelli1989} in the ultraviolet and infrared.

More recent studies set out to recalibrate the DIRBE/{\it IRAS}-derived dust map of SFD98 using the colour excesses of stars or standard cosmological sources from the Sloan Digital Sky Survey \citep[SDSS;][]{York2000}. \citet{Schlafly2010} and Schlafly \& Finkbeiner (2011)\nocite{Schlafly2011} used the colour excesses of $\sim 260,000$ stars and claimed that SFD98 overpredicted the Galactic reddening by up to $14$\%. Peek \& Graves 2010\nocite{Peek2010} used $\sim150,000$ passive galaxies in SDSS selected using upper limits on their $H\alpha$ and $[OII]$ equivalent widths to derive a new extinction map. They stated that SFD98 underestimated the reddening estimates at high latitudes up to $10-15$\%. \citet{Schlafly2014} derived a map of dust reddening for a large fraction of the northern hemisphere at Dec. $> 30^{\circ}$ from the photometry of half a billion stars from the Panoramic Survey Telescope and Rapid Response System 1 (Panstarrs, PS1). All these studies favoured the Milky Way extinction law from Fitzpatrick (1999; F99 hereafter)\nocite{Fitzpatrick1999}.
M{\"o}rtsell et al.~(2013)\nocite{Mortsell2013} also attempted to evaluate the scaling factor $p$ from the colour excesses of a variety of extragalactic sources such as quasars, BCG and red galaxies while simultaneously refining F99.

More recently, a higher resolution all-sky dust map was derived from a combination of the $5\arcmin$-resolution {\it Planck} data and the IRAS $100\mu$m data ({\it Planck} Collaboration XI 2014; P14 hereafter)\nocite{Planck2014}. P14 evaluated a dust-to-reddening conversion factor from the colour excesses of $\sim 53,000$ SDSS quasars. They adopted an F99 reddening curve and favoured a linear rescaling of their point-source-removed thermal dust radiance map $R$ (in units of W m$^{-2}$ sr$^{-1}$) over the dust optical depth at $353$~GHz ($\tau_{\rm 353}$) map to estimate reddening. They showed in particular that $R$ seems less affected by the cosmic infrared background anisotropies and that it correlates with $N_{\rm H}$ over a wider range of column density than $\tau_{\rm 353}$.

The methodologies used to derive reddening maps evidently differ from one study to another, for instance in the adopted prescription of the extinction law of the Milky Way or in the initial dust maps used for the calibration etc.; it would therefore be difficult to pin down all possible origins of the discrepancies between the existing reddening maps. In this paper however, we specifically investigate an inaccuracy of these works introduced by the systematic neglected fact that the source photometry is mainly obtained through broad-band filters and that, in consequence, the Galactic extinction does depend on the intrinsic spectral energy distribution (SED) of the observed source. 

The paper is organised as follows. Section 2 summarises the range of reddening expected along the line of sight of the {\it Euclid} wide field survey. Section 3 presents a number of caveats introduced by the failure to take into consideration the impact of the dependence of Galactic extinction with source SED. We first introduce the band-pass corrections required to renormalise the Milky Way absorption law for a given SED. As a proof of concept, we test the SED dependence of dust-to-reddening scaling factors using luminous red galaxies at $z < 0.4$ as standard crayons. We then quantify the impact of the source SED on photometric corrections. We then assess in section 4 how inaccurate Galactic extinction corrections lead to bias in source photometric redshift estimates. We summarise our findings in section 5. Appendix A introduces a prescription of Galactic extinction as a possible improved recipe for template-fitting codes, in particular for the {\it Euclid} template-fitting code Phosphoros. Appendix B investigates the dependence of the Galactic extinction with the uncertainties on extinction law prescription, in particular the uncertainties on the characteristic total-to-selective extinction $R_{\rm V}$. Appendix C confronts our analysis to the textbook calibration work of SFD98.

\section{Galactic extinction towards the {\it Euclid} wide survey}

\begin{figure*}
\begin{center}
\includegraphics[width=12cm,bb= 10 40 560 450]{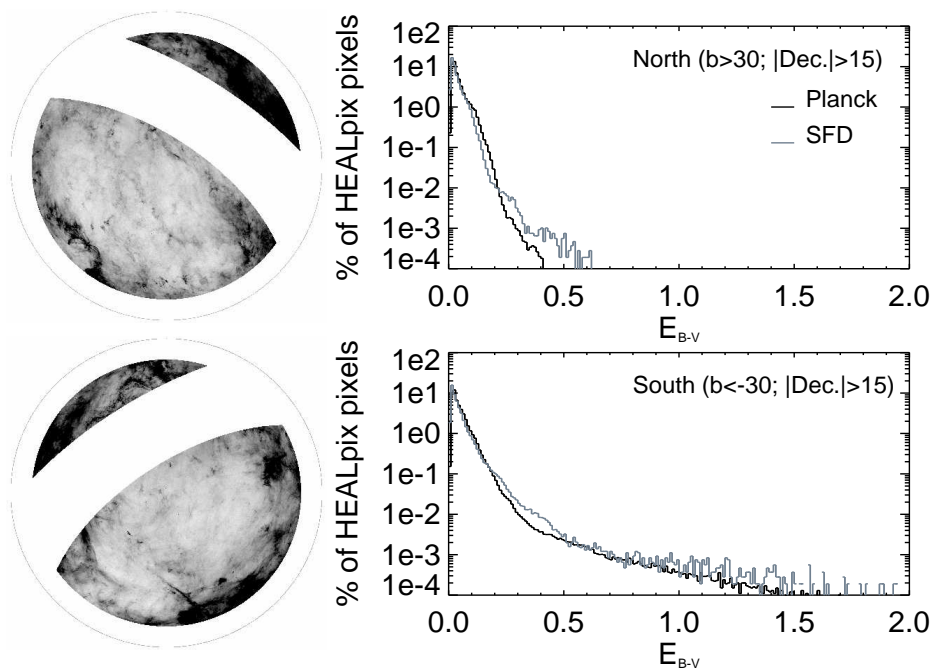}
\end{center}
\caption{Values of Galactic reddening along the line of sight of the {\it Euclid} wide survey northern (top) and southern Galactic hemisphere (bottom) footprints as defined in the {\it Euclid} Mission definition report. {\it Left:} Polar view of the {\it Planck} $E_{\rm B-V}$ maps (P14) with the Galactic plane ($|b| < 30$~deg) and ecliptic plane ($|$Dec.$| < 15^{\circ}$) removed. {\it Right:} Distribution (in \% of pixels) of $E_{\rm B-V}$ values for each Galactic hemisphere in P14 (black histogram) and SFD98 HEALPix $E_{\rm B-V} $maps (grey histogram).}
\label{fov}
\end{figure*}

According to the {\it Euclid} initial mission definition report, the so-called Red Book \citep{Laureijs2011}, the {\it Euclid} wide survey will cover a total of $15,000$ square degrees in the northern and southern Galactic hemispheres, preferentially at high Galactic latitude ($|b| > 30$~deg). The ecliptic plane will also be avoided by observing at $|$Dec.$| > 15^{\circ}$. Figure~\ref{fov} shows the $E_{\rm B-V}$ {\it Planck} maps and distribution of Galactic reddening in the foreseen {\it Euclid} wide survey footprint. The values range from $E_{\rm B-V} = 0$ to $\sim0.5$ in the northern Galactic hemisphere and up to $E_{\rm B-V} \sim 13$ (along the line of sight of the Large Magellanic Cloud) in the southern Galactic hemisphere. The final survey area of the {\it Euclid} wide survey has not been finalised at the time of writing, although the priorities regarding the survey strategy will favour regions of Galactic extinction $E_{\rm B-V} < 0.3$. Due to visibility issues however, regions with higher $E_{\rm B-V}$ will also be observed ({\it Euclid} Consortium Survey Working Group; private communication). 

In the northern Galactic hemisphere within the {\it Euclid} wide survey footprint mentioned above, regions with $E_{\rm B-V} > 0.3$ account for $\sim0.01$\% according to P14 ($\sim0.04$\% for SFD98) $E_{\rm B-V}$ maps. Regions of such high extinction are more commonly found in the southern Galactic hemisphere and account for $0.3$\% (P14; $\sim0.5$\% for SFD98). Regions with lower but still significant reddening ($E_{\rm B-V} > 0.1$) account for $\sim7$\%  (P14; $\sim3$\% for SFD98) in the northern Galactic hemisphere and $\sim8$\% (P14; $7$\% for SFD98) in the southern Galactic hemisphere.

\section{Dependence of reddening estimates with SED}

\subsection{Definitions}

The total extinction in a filter $X$, $A_{\rm X}$ is equal to $-2.5$ log$(f_{\rm obs,X} / f_{\rm int,X})$ where $f_{\rm int,X}$ the intrinsic flux density of a source measured within a filter $X$ and $f_{\rm obs,X}$ its observed flux density affected by Galactic extinction measured within the same filter, expressed respectively as:

\begin{equation}
f_{\rm int,X} = \frac{\int_{\rm X} f_{\rm sed}(\lambda) F_{\rm X}(\lambda) d\lambda}{\int_{\rm X} F_{\rm X}(\lambda) \frac{c}{\lambda^2} d\lambda} ,
\end{equation}

\begin{equation}
f_{\rm obs,X} = \frac{\int_{\rm X} f_{\rm sed}(\lambda) 10^{-0.4 A_{\rm \lambda}} F_{\rm X}(\lambda) d\lambda}{\int_{\rm X} F_{\rm X}(\lambda) \frac{c}{\lambda^2} d\lambda} .
\end{equation}

Here, $f_{\rm sed}(\lambda)$ is the source SED and $F_{\rm X}(\lambda)$ the transmission curve of the filter. The denominator factor compensates for the fact that the integration within the filter is done in wavelength instead of frequency. $A_{\rm \lambda}$ is the extinction of the Milky Way at wavelength $\lambda$ usually expressed as $A_{\rm \lambda} = E_{\rm B-V} \, k_{\rm \lambda}$ where $k_{\rm \lambda}$ the absorption law normalised to $E_{\rm B-V}$. If the observed fluxes are obtained through broad-band filters, the Galactic reddening of a source therefore depends of its intrinsic SED and in particular of its shape within the filter. 

Extragalactic surveys commonly adopt an estimate of Galactic extinction in a filter $X$ given by $A_{\rm X} = E_{\rm B-V} \, k_{\rm pivot}$. $E_{\rm B-V}$ is the value of the reddening along the line of sight of the studied source (or in some studies simply the reddening value at the centre of the extragalactic field) and $k_{\rm pivot}$ a fixed value of the Milky Way absorption law usually adopted at the pivot\footnote[1]{In the rest of the paper, we adopt the following definition for a filter pivot wavelength: $\lambda_{\rm pivot,X} = \sqrt{\int \lambda F_{\rm X} d\lambda / \int F_{\rm X} d\lambda / \lambda}$ \citep{Tokunaga2005}.} wavelength of the filter. This approach is for instance implemented within the widely used online NASA Extragalactic Database (NED) Galactic extinction calculator\footnote[2]{http://irsa.ipac.caltech.edu/applications/DUST/}. The NED calculator makes use of the Schlafly \& Finkbeiner (2011) recalibration of SFD98 reddening map mentioned in section 1. 

The dependence of Galactic extinction with SED is disregarded in wide field extragalactic studies since the nature of a source and thus its SED shape is not known a-priori. The impact of SED on Galactic extinction is also particularly significant in regions of high Galactic extinction (large $E_{\rm B-V}$ along the line of sight) that are classically avoided by extragalactic wide-field surveys. For reference, the ranges of Galactic reddening along the line of sight of the five well studied extragalactic fields of the CANDELS survey are $0.009-0.015$ for GOODS-south, $0.014-0.018$ for GOODS-north, $0.010-0.019$ for EGS, $0.020-0.030$ for COSMOS and $0.024-0.032$ for UDS. These values are extracted from the {\it Planck} $E_{\rm B-V}$ map. The amount of Galactic extinction and therefore its dependence with source SED is directly related to the shape of the Milky Way absorption curve, that is to say mainly a decreasing slope with $\lambda$ and is thus much more severe in the optical wavelengths. 

\subsection{Band-pass corrections}

For the rest of the analysis, we adopt the parametrisation of the Milky Way absorption law from Fitzpatrick (1999) assuming $R_{\rm V} = 3.1$ corresponding to the mean value of $R_{\rm V}$ for the diffuse interstellar medium. Appendix B explores the impact of the uncertainties of $R_{\rm V}$ on galactic extinction and photometric redshift estimates.

The F99 extinction law was calibrated using the colour excesses of main sequence B5 stars. Fitzpatrick (1999) uses $B-$ and $V$-band measurements in the Johnson filter system; we therefore adopt the Johnson filter throughputs for the present analysis. In order to obtain a universal extinction curve, independent of the amount of extinction along the line-of sight of the stars used for the calibration, the extinction curve $A_{\rm \lambda}$ is normalised by $E_{\rm B-V}$.

A consequence of the use of the broad-band $B$ and $V$ filters for the calibration of F99 is that for other SED than B5 stars, the observed value of the reddening $E_{\rm B-V} = A_{\rm B} - A_{\rm V}$, namely the difference between the total extinction in the $B$-band and $V$-band (see equation 3), will be different from the $E_{\rm B-V}$ used to normalise the absorption law $k_{\rm \lambda} = A_{\rm \lambda} / E_{\rm B-V}$. This caveat is often mentioned in the literature (for instance in P14, Appendix E) but these works do not quantify its effective impact. Band-pass corrections are thus required to adequately re-normalise the Milky Way absorption law for sources with different SEDs.

Before moving forward, as confusion may arise from the use of $B$ and $V$ filter terminology, we restrict the use of the $E_{\rm B-V}$ notation to the observed reddening i.e,~the difference between total extinction in the $B$- and $V$-band. We use the nomenclature $A_{\rm \lambda} = p_{sed}D k_{\rm \lambda}$ in equation 2, where the amount of extinction is assumed to be a linear rescaling (by a scaling factor $p_{sed}$) of the Galactic dust value along the line of sight ($D$) (see section 1). For a source of a given SED, 

\begin{equation}
E_{\rm B-V}^{sed} = -2.5 log\left( \frac{\int_{\rm B} f_{\rm sed} 10^{-0.4 p_{\rm sed} D k_{\rm \lambda}} F_{\rm B} d\lambda}{\int_{\rm V} f_{\rm sed} 10^{-0.4 p_{\rm sed} D k_{\rm \lambda}} F_{\rm V} d\lambda} \, \frac{\int_{\rm V} f_{\rm sed} F_{\rm V} d\lambda}{\int_{\rm B} f_{\rm sed} F_{\rm B} d\lambda} \right).
\end{equation}

We restate that for a SED of a B5 star by construction,

\begin{equation}
E_{\rm B-V}^{B5} = p_{\rm B5}D.
\end{equation}

For a source with any other SED however, the derived $E_{\rm B-V}^{sed}$ will differ from $p_{\rm B5}D$. For example, if we derive the $E_{\rm B-V}$ from the SED of an elliptical galaxy (we adopt the {\rm LePhare} COSMOS `Ell1\_A\_0.sed' template) at $z = 0.1$ and a $pD = 0.1$ along the line of sight, we obtain $E_{\rm B-V}^{sed} = 0.09$ and not $0.1$. In order to recover $E_{\rm B-V}^{sed} = pD$ for a given SED, we need to renormalise the absorption law by a band-pass correction following 

\begin{equation} 
p_{\rm sed}D \; k_{\rm \lambda}= p_{\rm B5}D \; \frac{k_{\rm \lambda}}{{\rm bpc}_{\rm sed}}.
\end{equation}

bpc$_{\rm sed}$ designates the SED-dependent band-pass correction required to convert an extinction law calibrated with B5 stars to another SED. Although the band-pass correction technically has to be applied to the extinction law, we will assume in the rest of the paper that the definition of $k_{\rm \lambda}$ is fixed (to F99) and apply the band-pass correction on the scaling factor $p$ with $p_{\rm sed} = p_{\rm B5} / {\rm bpc}_{\rm sed}$.

\begin{figure}
\begin{center}
\includegraphics[width=8cm,bb = 0 30 520 420]{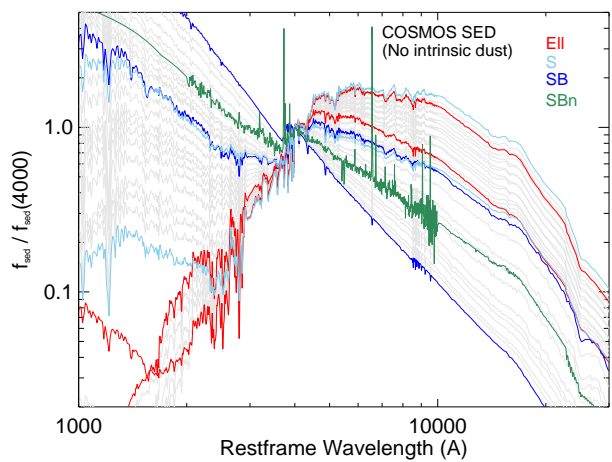}
\end{center}
\caption{The {\rm LePhare} COSMOS template library which includes passive (`Ell'), star-forming (`S0' to `Sd' types) and so-called starburst templates (`SB' i.e.~galaxies with young stellar population). For the sake of clarity, only the most extreme templates that encompass the whole flux versus wavelength space covered by a given spectral type, are coloured (`Ell' in red, `S' in light blue and `SB' in blue). A {\rm LePhare} CFHTLS template with added nebular emission (`SBn') is also plotted in green. We expand the full library of templates by dust-reddening and redshifting the non-dusty restframe templates using a grid of intrinsic dust extinction and redshift.} 
\label{sed}
\end{figure}

If we consider again the case example of an elliptical galaxy at $z = 0.1$ and a value $p_{\rm B5}D = 0.1$, assuming a band-pass correction bpc$_{\rm sed} = 0.9$ (i.e.~$p_{\rm sed}D \sim 0.11$) would lead to $E_{\rm B-V}^{sed} = 0.1$. The band-pass corrections for each template can therefore be computed using

\begin{equation} 
{\rm bpc}_{\rm sed} = \frac{E_{\rm B-V}^{sed}}{E_{\rm B-V}^{B5}} ,
\end{equation}

\noindent following equations 3 and 4. We explore the range of band-pass corrections potentially encountered in extragalactic studies by deriving bpc$_{\rm sed}$ for galaxies with a range of intrinsic SEDs. We exploit the so-called COSMOS galaxy template library distributed with the {\rm LePhare} software\footnote[3]{http://www.cfht.hawaii.edu/~arnouts/LEPHARE/lephare.html} that performs photometric redshift estimates using template-fitting techniques. The COSMOS library comprises seven passive, $12$ S0 to Sd and $12$ starburst templates. Figure~\ref{sed} shows the $31$ COSMOS templates. The template library is expanded by dust-reddening and redshifting the non-dusty templates using a grid of redshift $z = [0 - 2]$ (by bins of $\Delta z = 0.01$) and a grid of intrinsic dust extinction\footnote[4]{The reddening by the dust contained within the galaxy itself is designated by $E_{\rm B-V}^{Int}$, to be distinguished from $E_{\rm B-V}$, the absorption by the Milky Way along the line of sight.} $E_{\rm B-V}^{Int} = [0 - 0.3]$ (by bins of $\Delta E_{\rm B-V}^{Int} = 0.1$) adopting a dust extinction law from \citet{Calzetti2000} that is commonly adopted for absorption by intrinsic dust in extragalactic sources (see for example Ilbert et al. 2009, Hildebrandt et al.~2010 but see also Kriek et al.~2013). \nocite{Ilbert2009, Hildebrandt2010, Kriek2013}

\begin{figure*}
\begin{center}
\includegraphics[width = 14cm,bb = 20 20 530 310]{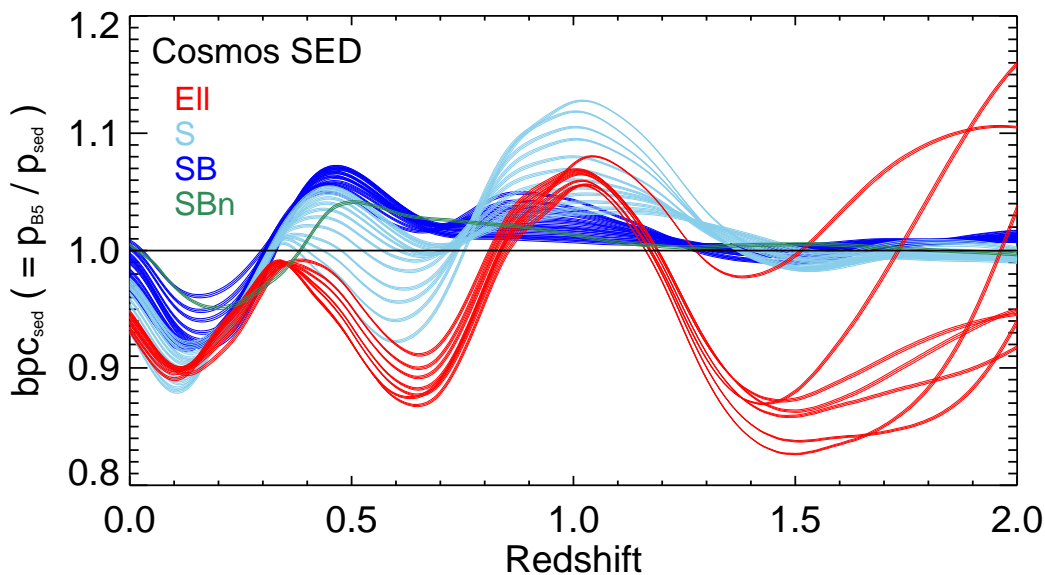}
\end{center}
\caption{Band-pass corrections for the COSMOS (`Ell' in red, `S0' to `Sd' in light blue and `SB' in blue) and CFHLTS SBn (green) galaxy templates in function of redshift (from $z = 0$ to $2$) for models with no intrinsic dust ($E_{\rm B-V}^{Int} = 0$). The dependence of the band-pass correction with the $pD$ intensity itself is shown for each template by a set of three lines corresponding to $0.1 < p_{\rm B5}D < 0.3$.}
\label{conversmodels}
\end{figure*}

Figure~\ref{conversmodels} shows examples of band-pass corrections for the different templates as a function of their redshift (from $z = 0$ to $2$) for models with no applied intrinsic dust ($E_{\rm B-V}^{Int} = 0$). We mention a few particular cases in the following for illustration purposes. Star-forming galaxies at $z > 1.5$ have bpc$_{\rm sed} \sim 1$ which means that no correction is required to renormalise the Milky Way absorption law. A passive galaxy at $z = 1$ would however require a bpc$_{\rm sed} \sim 1.05$, meaning that the $p_{\rm sed}D$ needed to renormalise the F99 Milky Way absorption law for this type of source is $5$\% higher than $p_{\rm B5}D$. The presence of emission lines (green model) does not seem to have a major impact on band-pass corrections. 

As expected, the band-pass correction values are governed by the SED characteristic features within the $B$ and $V$ filters. For example, the increase in band-pass correction systematically observed between $z = 0.1$ and $z = 0.3$ for all sources, independently of their type, is due to the $4000$\AA~break passing through the $B$ and $V$ filters at these redshifts. 

The value $pD$ itself also has an impact on the shape of the spectral energy distribution of a source and therefore on the required band-pass correction. In Figure~\ref{conversmodels}, the band-pass corrections for a $p_{\rm B5}D$ along the line of sight varying from $0.1$ to $0.3$ are shown by the thickness of the lines. Since $A_{\rm X}$ (and therefore $E^{sed}_{\rm B-V}$) scales almost linearly with $pD$ within the investigated regime ($p_{\rm B5}D < 0.3$; see Appendix A), the dependence of the band-pass corrections with $pD$ intensity is small ($\sim 0.2$\% at most). We therefore neglect this effect in the rest of the analysis and adopt $p_{\rm B5}D = 0.1$ as our standard value to derive band-pass corrections.

The most extreme (minimum and maximum) band-pass corrections of the whole sample of models are of the order of $\sim20$\% ($0.8 \leq {\rm bpc}_{\rm sed} \leq 1.2$). Neglecting these significant corrections leads to a commonly overlooked inaccuracy in the procedures classically used to derive the dust-to-reddening scaling factor $p$ while calibrating Galactic reddening maps. 

\begin{figure}
\begin{center}
\includegraphics[width = 8.4cm,bb = 20 30 450 300]{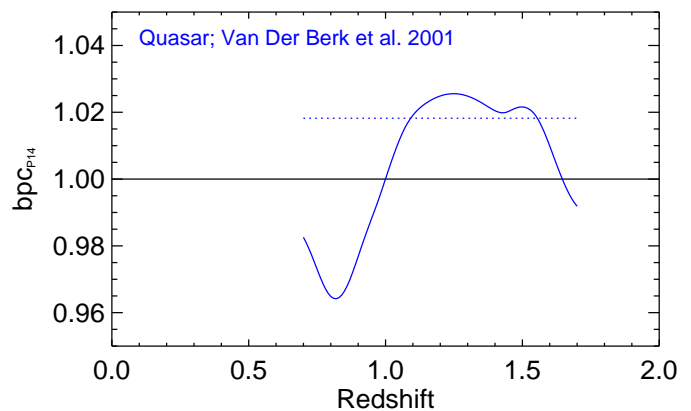}
\end{center}
\caption{Band-pass corrections derived from the composite quasar spectrum of \citet{VandenBerk2001}. This composite spectrum was used in P14 to estimate the reddening along the line of sight of a quasar calibration sample at $0.7 < z < 1.7$. The dotted line marks the median band-pass correction of the sample corresponding to bpc$_{\rm P14} = 1.018$.}
\label{converstempquasar}
\end{figure}

\subsection{The calibrations of reddening maps are SED-dependent}

The dust-to-reddening scaling factor $p$ is classically derived from the estimates of the colour excess of a sample of stars or standard cosmological sources with well-known SED such as quasars, luminous red galaxies etc and known redshift. The observed colours of each source are compared to the predicted colours of a template adopted to be representative of the source SED, redshifted at the redshift of the sources and to which a range of Galactic reddening values are applied. The value of the reddening that optimises the match, through, for example, $\chi^2$ minimisation, between observed and reddened modelled colours is adopted as the reddening estimate along the line of sight and confronted to the value of the dust map at the source sky position. 

The scaling factor $p$ is then derived by linearly correlating dust and reddening for large samples of sources. For an adopted Galactic extinction law and if band-pass corrections are not considered, the derived scaling factor $p$ will depend on the source SED used for the calibration. In other words, if the calibration is repeated adopting the same extinction law and the dust map but using a sample of sources with different SEDs, the derived value of $p$ will differ. The expected difference in scaling factors can be predicted by the band-pass corrections. If $p$ is calibrated using two samples of sources with representative spectral energy distributions `sed1' and `sed2', then, following equation 5, $p_{\rm sed1} = p_{\rm B5} / {\rm bpc}_{\rm sed1}$ and $p_{\rm sed2} = p_{\rm B5} / {\rm bpc}_{\rm sed2}$ i.e.~

\begin{equation}
p_{\rm sed2} = \frac{p_{\rm sed1} \, {\rm bpc}_{\rm sed1}}{{\rm bpc}_{\rm sed2}}.
\end{equation}

For example, P14 used the colour excesses of SDSS quasars to linearly rescale the {\it Planck} point-source-removed thermal dust radiance map $R$ to an $E_{\rm B-V}$ reddening map. They considered quasars from the final release of the SDSS-II quasar catalogue \citep{Schneider2010} and limited their calibration sample to sources with $0.7 < z < 1.7$ where the $Ly\alpha$ emission line does not fall in the $ugriz$ filters. An F99 Milky Way absorption law was adopted. They compared the quasar observed colours to colours derived from the composite quasar spectrum of \citet{VandenBerk2001}, then estimated the reddening along the line of sight of each source and found $p_{\rm P14} = E_{\rm B-V} / R = (5.40 \pm 0.09) \times 10^5$~m$^2$ sr W$^{-1}$. Since P14 used quasars with a range of redshifts and therefore a range of different SED, $p_{\rm P14}$ is an `average' scaling factor over their quasar sample. 

Figure~\ref{converstempquasar} shows the range of band-pass corrections derived for the Vanden Berk et al.~2001 spectrum in the redshift interval $0.7 < z < 1.7$. We re-extract the quasars within that redshift range from the \citet{Schneider2010} catalogue and derive the band-pass correction of each quasar at its given redshift using the \citet{VandenBerk2001} spectrum. The median of the corrections is bpc$_{\rm P14} = 1.018$ (Figure~\ref{converstempquasar}, blue dotted line; $2/3$ of their sample is at $z > 1.1$) which indicates that if B5 stars had been used to calibrate the radiance map instead of the quasar sample from Schneider et al.~2010, the analysis would have converged to $p_{\rm B5} \sim 5.5$~m$^2$ sr W$^{-1}$ ($\sim 5.40 \times 1.018$). We note that this value of $p_{\rm B5}$ corresponds to the scaling factor we would derive from the colour excesses of B5 stars if these stars were all outside the Milky Way, in other words if they would trace the full Galactic absorption along the line of sight.

\begin{figure}
\begin{center}
\includegraphics[width = 9cm,bb = 25 35 555 595]{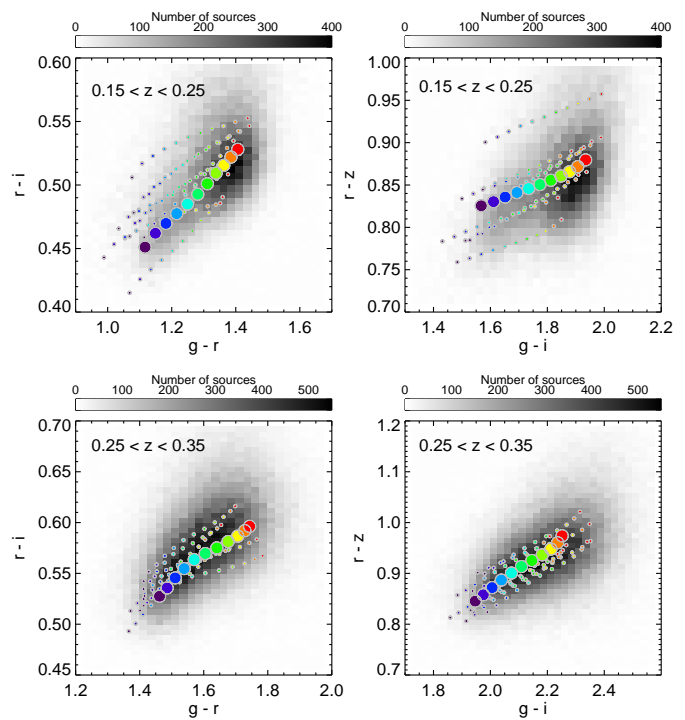}
\end{center}
\caption{Colour-colour diagrams ({\it left panels:} $r - i$ vs. $g - r$; {\it right:} $r - z$ vs. $g - i$) of SDSS LRG in two redshift bins ({\it top:} $0.15 < z < 0.25$; {\it bottom:} $0.25 < z < 0.35$). The grey colour map accounts for the density of sources. Galaxy colours are derived from extinction-corrected $griz$ model magnitudes (see text for details). The coloured small-dotted tracks show the colours of two template libraries from Greisel et al.~(2013) for LRG at $z \sim 0.2$ and $z \sim 0.3$ (top: $z = [0.15,0.25]$; bottom: $z = [0.25,0.35]$; bin $= 0.01$; purple to red shows increasing redshift). The two optimal templates at $z \sim 0.2$ and $z \sim 0.3$ adopted as reference template for the calibration of the dust-to-reddening factor are shown by the larger dotted tracks.}
\label{colourcolour}
\end{figure}

\begin{figure*}
\begin{center}
\includegraphics[width = 18.6cm,bb = 5 240 585 465]{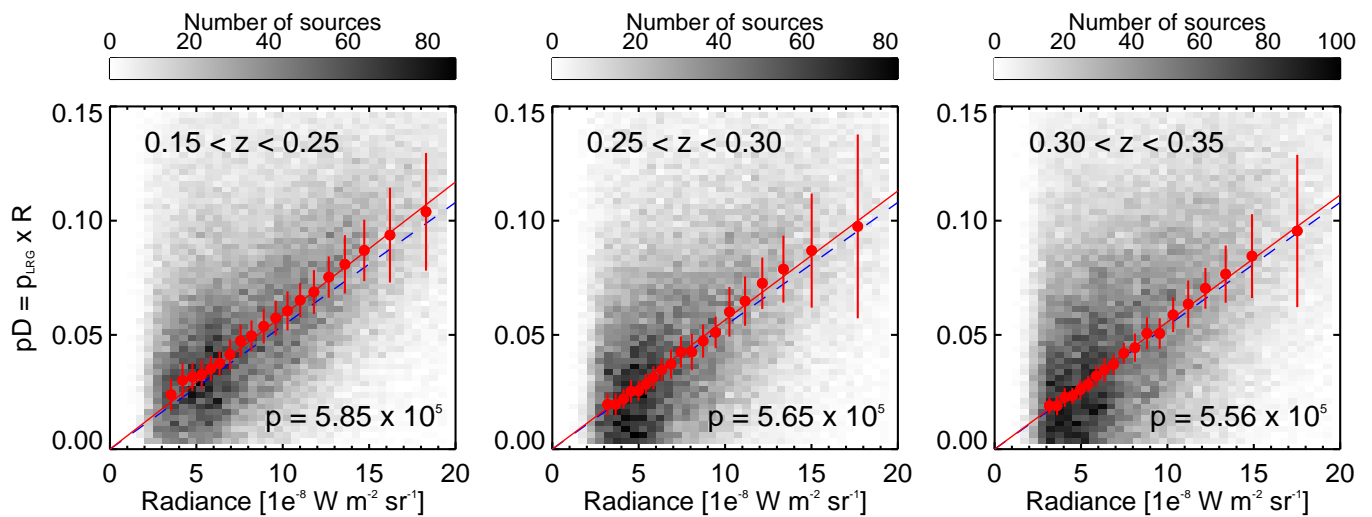}
\end{center}
\caption{Estimates of $pD$ as a function of $R$ from the {\it Planck} radiance map measured along the line of sight of SDSS red galaxies at $0.15 < z < 0.25$ (left), $0.25 < z < 0.30$ (middle) and $0.30 < z  < 0.35$ (right). The grey colour map accounts for the density of sources. The red dots (and error bars) represent the median (and standard deviation) estimated through bootstrapping resampling of $pD$ and $R$ within the range $2 < R < 20 \times 10^{-8}$~W m$^{-2}$ sr$^{-1}$ split in $20$~bins of equal number of sources. The red line is a linear fit to the dots. We find $p_{\rm LRG} \sim 5.85$, $5.65$ and $5.56 \times 10^5$~m$^2$ sr W$^{-1}$ for $0.15 < z < 0.25$, $0.25 < z < 0.30$ and $0.30 < z < 0.35$ respectively. The blue dashed line shows the $p_{\rm P14} = 5.4 \times 10^5$ derived by P14 from a quasar sample at $0.7 < z < 1.7$ (P14).}
\label{boss}
\end{figure*}

\subsection{Testing the SED dependence of $p$ using red galaxies}

We further test the impact of a source SED on the calibration of the scaling factor by deriving new estimates of $p$ from the colour excesses of $z < 0.4$ luminous red galaxies (LRG). 

The LRG were selected from the SDSS-III Baryon Oscillation Spectroscopic Survey (BOSS) by means of colour cuts following \citet{Padmanabhan2005} and flagged as such in the SDSS database\footnote[5]{The sql request includes $(BOSS\_TARGET1 \& 1) != 0$, $SPECPRIMARY == 1$, $ZWARNING\_NOQSO == 0$ and $TILEID >= 10324$ following https://www.sdss3.org/dr9/algorithms/boss\_galaxy\_ts.php}. Only sources with an accurate spectroscopic redshift (i.e.~$\Delta z_{\rm spec} \leq 0.002$) were selected. We retrieved the LRG non extinction-corrected $griz$ model photometry from the SDSS Catalog Archive Server. Model fluxes have been shown to provide the best estimates for colours\footnote[6]{www.sdss.org/dr12/algorithms/magnitudes/}. We do not use the $u$-band photometry due to the large errors in SDSS $u$-band measurements \citep{Padmanabhan2008} as well as the large scatter in colours using $u$-band, classically attributed to variations in stellar population. 

The reddening is estimated by confronting the LRG observed colours with the predicted colours of a red galaxy spectrum reddened by a grid of $pD$ value. Figure~\ref{colourcolour} shows the colour distribution of LRG in $griz$ colour-colour diagrams. In order to choose the reference template for our LRG sample (and only at this stage), the SDSS fluxes are corrected from Galactic extinction by a first-order correction using the reddening along the line of sight of each source from P14 multiplied by the F99 reddening curve value at the corresponding SDSS filter pivot wavelength (see Figure~\ref{colourcolour}). We make use of the library of red galaxy templates from \citet{Greisel2013} (G13, hereafter) optimised for LRG at $z < 0.5$ per bin of redshift i.e.~for $z \sim 0.02$, $0.1$, $0.2$, $0.3$ and $0.4$. G13 generated their template libraries via a superposition of Bruzual \& Charlot (2003)\nocite{Bruzual2003} composite stellar population and burst models and isolated a subsample of SEDs that adequately matches the colour space of LRG at a given redshift bin. We split the SDSS LRG sample in two redshift bins, $0.15 \leq z < 0.25$ ($\sim 54,000$ sources) and $0.25 \leq z \leq 0.35$ ($\sim 118,000$ sources) and choose for each redshift bin one representative template from the G13 LRG libraries at $z \sim 0.2$ and $z \sim 0.3$ respectively. Figure~\ref{colourcolour} shows the G13 sets of templates (N. Greisel; private communication) at $z \sim 0.2$ ($12$ models) and $z \sim 0.3$ ($10$ models). The best-fit template for each source is chosen through $\chi^2$ minimisation in $griz$ from the $z \sim 0.2$ templates for sources at $0.15 \leq z < 0.25$ and from the $10$ $z \sim 0.3$ templates for sources at $0.25 \leq z \leq 0.35$, redshifted to the LRG spectroscopic redshift. We then adopt the template that fits the majority of galaxies at each redshift bin. As expected, the chosen templates have colours consistent with the core of the LRG distribution in Figure~\ref{colourcolour} (large dots).

We then compare the observed $griz$ SDSS photometry of each LRG to the colours of the reference template redshifted to the LRG spectroscopic redshift to which we apply a grid of $pD$ following $-0.10 < pD < 0.20$ with $\Delta(pD) = 0.0001$. The range of sampled $pD$ was extended below zero in order not to bias the estimate of $pD$ (see for example Noll et al.~2009, figure~8)\footnote[7]{If the distribution of $pD$ is Gaussian-shaped and extends below $pD < 0$, the limitation of the parameter space to $pD > 0$ will bias the median probability of the posterior distribution towards $pD > 0$}\nocite{Noll2009}. 

The best reddened fit, chosen through $\chi^2$ minimisation, procures the estimation of $pD$ along the line of sight of the source. We then confront $pD$ with the value of the {\it Planck} radiance map (chosen as our reference `dust' map) at the position of the source. As already mentioned in previous calibration works (for instance in P14), given the small range of reddening considered (mainly $pD < 0.1$), the large number of sources allows to average over both SDSS photometric errors and the discrepancies between the LRG real colours and the adopted `optimal' reference template colours. 

Figure~\ref{boss} shows the correlation of the $pD$ estimates with {\it Planck} $R$. Due to the available large statistics, the $z \sim 0.3$ redshift bin is split in two: $0.25 \leq z < 0.30$ and $0.30 \leq z \leq 0.35$. We consider sources in the radiance range $2 < R < 20$~W m$^{-2}$ sr$^{-1}$. This range is split in $20$~bins such that all bins contain the same number of sources ($\sim 2000-2500$). The median $pD$ and $R$ and associated dispersion are derived using bootstrap resampling; we draw from the $pD$ and $R$ distribution in each radiance bin a new sample of values of the same number of sources, repeat the process $1000$~times and compute its median and standard deviation. Figure~\ref{boss} shows the median $(R;pD)$ per radiance bin. We then use a linear regression to derive the scaling factor $p$ for the different redshift bins, normalising the fit such that zero dust column density corresponds to zero extinction and find

\begin{figure}
\begin{center}
\includegraphics[width = 8cm,bb = 30 80 480 480]{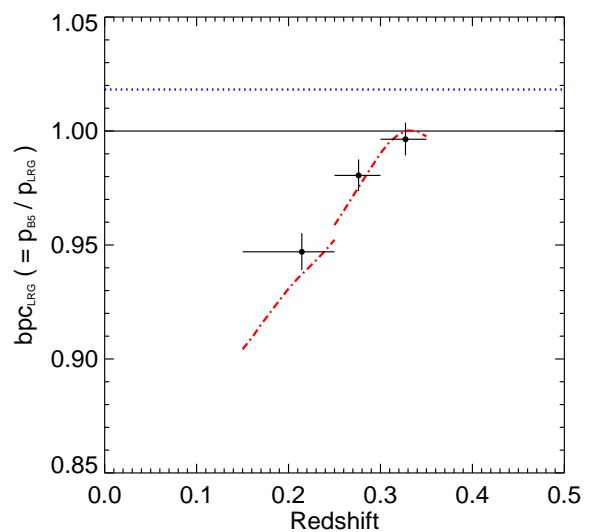}
\end{center}
\caption{Band-pass corrections derived from two red galaxy templates (dotted dashed red lines) from Greisel et al.~(2013). These templates are adopted as standard reference SED for the LRG sample used to derive the scaling factors $p_{\rm LRG}$ that convert the {\it Planck} radiance map to reddening values (see section 3.4). The first is used to calibrate the $z = [0.15-0.25]$ sample, the second to calibrate both the $z = [0.25-0.3]$ and $z = [0.3-0.35]$ samples. The $p_{\rm LRG}$ obtained for the three samples are shown by the black dots with the plotted values corresponding to bpc$_{\rm LRG} = p_{\rm B5} / p_{\rm LRG}$ where the optimal $p_{\rm B5}$ is found to be equal to $5.54 \times 10^5$~m$^2$ sr W$^{-1}$. The median band-pass correction bpc$_{\rm P14} = 1.018$ (see Figure~\ref{converstempquasar}) is also plotted in blue for reference.}
\label{converstemp}
\end{figure}

\begin{figure*}
\begin{center}
\includegraphics[width=13cm,bb = 5 40 460 420]{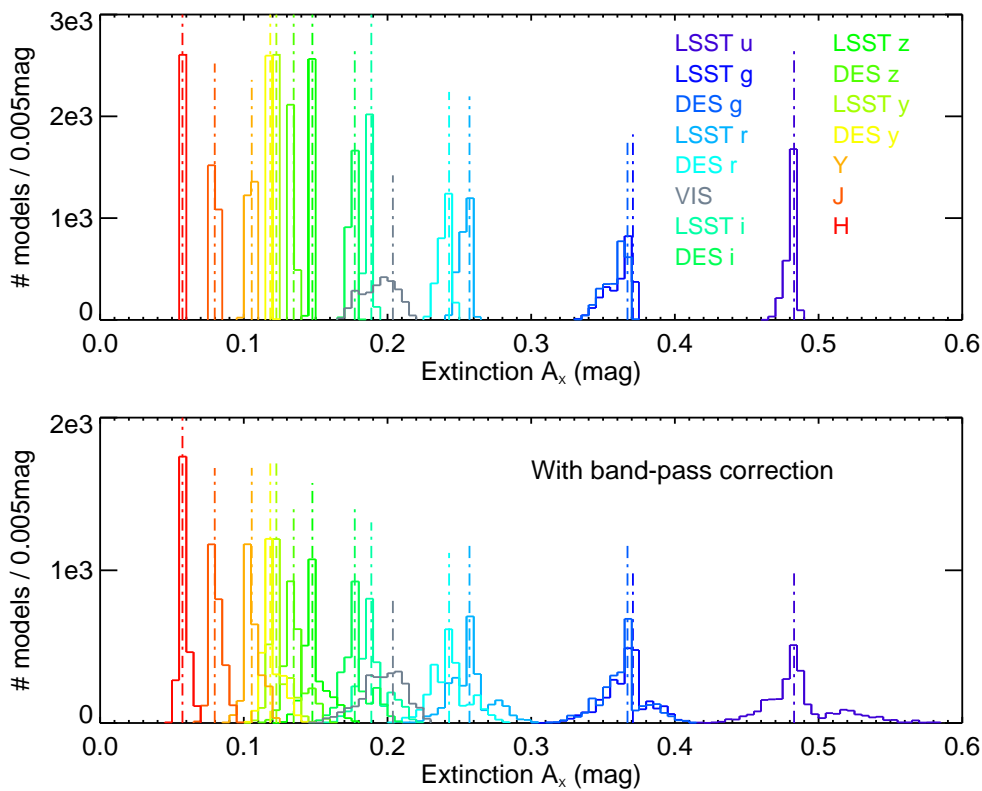}
\end{center}
\caption{Distribution of Galactic extinction $A_{\rm X}$ (in magnitude) for $pD = 0.1$ for a library of $2604$ extragalactic source SEDs and a set of optical to near-infrared filters ($0.005$~mag bin size, purple to red for redder filters). In both panels, the vertical lines indicate the classically adopted photometric correction obtained by multiplying the reddening along the line of sight with a fixed value of the reddening curve at the filter pivot wavelength. The top panel shows the distribution of $A_{\rm X}$ (equation 1 and 2) assuming $A_{\rm \lambda} = 0.1 \, k_{\rm \lambda}$ where band-pass corrections are neglected. The bottom panel shows the distribution of Galactic extinction, this time re-normalising first the absorption law adequately by applying the suited SED band-pass correction following $A_{\rm \lambda} = 0.1 \, k_{\rm \lambda} / {\rm bpc}_{\rm sed}$.}
\label{deltam_allps}
\end{figure*}

\begin{itemize} 
\item $p_{\rm LRG[0.15-0.25]} = (5.85 \pm 0.05) \times 10^5$~m$^2$ sr W$^{-1}$, 
\item $p_{\rm LRG[0.25-0.30]} = (5.65 \pm 0.04) \times 10^5$~m$^2$ sr W$^{-1}$,
\item $p_{\rm LRG[0.30-0.35]} = (5.56 \pm 0.04) \times 10^5$~m$^2$ sr W$^{-1}$. 
\end{itemize}

We check that the scaling factor estimates and their relative ratios are consistent with what is expected from predictions of band-pass corrections. Figure~\ref{converstemp} shows the band-pass corrections bpc$_{\rm LRG}$ derived from the two LRG templates adopted for $z \sim 0.2$ and $z \sim 0.3$. The median redshifts of sources used for the estimate of the scaling factors (i.e.~with $2 < R < 20$) are $z \sim 0.22$, $0.28$ and $0.33$. At these redshifts, the predictions from the templates show that

\begin{itemize}
\item bpc$_{\rm LRG[0.15-0.25]} = 0.94$, 
\item bpc$_{\rm LRG[0.25-0.30]} = 0.98$, 
\item bpc$_{\rm LRG[0.30-0.35]} = 1.00$. 
\end{itemize}

From equation $6$, we see for instance that $p_{\rm LRG[0.25-0.30]}$ should be consistent with $p_{\rm LRG[0.30-0.35]} \, {\rm bpc}_{\rm LRG[0.30-0.35]} / {\rm bpc}_{\rm LRG[0.25-0.30]}$ and that indeed, the two quantities agree. The consistency check was carried out for all redshift bins. The estimates of the ratios between scaling factors estimates are fully in agreement with the predictions of models. Figure~\ref{converstemp} shows the three values of $p$ derived by this new calibration work. The plotted quantities correspond to bpc$_{\rm LRG} = p_{\rm B5} / p_{\rm LRG}$ where the normalisation factor $p_{\rm B5}$ is obtained using a weighted mean between model predictions (red lines) and scaling factors (black dots) following

\begin{equation}
p_{\rm B5} = \sum\limits_{\rm n=1}^{3} \frac{p_{\rm LRG} / {\rm bpc}_{\rm LRG}}{\sigma^{2}_{\rm p_{\rm LRG}}} / \sum\limits_{\rm n=1}^{3} \frac{(1/{\rm bpc}_{\rm LRG})^{2}}{\sigma^{2}_{\rm p_{\rm LRG}}}
\end{equation}

\noindent where $\sigma_{\rm p_{\rm LRG}}$ the errors on the $p_{\rm LRG}$ estimates and the sums are done over the three values corresponding to the three studied redshift intervals. We find $p_{\rm B5} = 5.54 \times 10^5$~m$^2$ sr W$^{-1}$, a value consistent with our first estimate of $p_{\rm B5}$ in section 3.3 derived from $p_{\rm P14}$ estimate. In Figure~\ref{converstemp}, the quantities are placed on the x axis at the median redshift of the LRG subsample used to calculate $p$. Following equation 5 again, $p_{\rm P14} = p_{\rm B5} / {\rm bpc}_{\rm P14} = 5.44 \times 10^5$~m$^2$ sr W$^{-1}$, a value that is also fully consistent with the P14 estimate of $(5.40 \pm 0.09) \times 10^5$~m$^2$ sr W$^{-1}$. We note however that, although it is not explicitly mentioned in P14, their figure~22 suggests that the authors did not impose their linear fit to originate from $(R,E_{\rm B-V}) = (0,0)$ in contrast to the present work.

The full analysis was repeated using the second-best template from the Greisel et al.~(2013) library to derive reddening along the line of sight of our LRG sample. The derived scaling factors using these alternative standard LRG SED were fully consistent with the expected band-pass corrections predicted by the models.

We also replicate the present analysis but this time, no longer by estimating the dust-to-reddening scaling factors from the P14 radiance map but from the DIRBE/{\it IRAS}-combined dust map that was used by SFD98 to derive their $E_{\rm B-V}$ map. The results are summarised in Appendix C that shows once again that the scaling factors derived for our sample of LRG for different redshift bins are consistent with predictions of template-derived band-pass corrections. We however raise the caveat brought about by the adopted extinction law prescription used for calibration (see also Appendix B).

\begin{table*}
\caption{Examples of Galactic extinction}
\label{tablegroup}
\centering
\begin{tabular}{l c | c c c c c}
\hline
Filter	&	Pivot		&	$A_{\rm X}$		&	$A_{\rm X}$		& $\Delta A_{\rm X}$ 		&	$A_{\rm X}$ 		& $\Delta A_{\rm X}$ 	\\
	&	nm		&	@pivot		&	median		& max-min			&	median		& max-min			\\
	&			&				&	SED 			& SED 				&	CANDELS 	& CANDELS 			\\
%	& 			&	$pD = 0.1$	&	$0.1$ 		& $0.1$ 				&	$0.1$ 		& $0.1$ 				\\
	& 			&	($1$)			&	($2$)			& ($3$)				&	($4$)			& ($5$)				\\
\hline 
LSST u	&	$358.3$	& $0.483$	 & 	$0.483$	&	$0.167$	&	$0.482$	&	$0.151$	\\%		$0.480$ &	$0.032$	& $0.483$ & $0.031$ \\
LSST g 	&	$477.0$	& $0.371$ & 	$0.368$	&	$0.108$	&	$0.369$	&	$0.092$	\\%	$0.362$ &	$0.051$ 	& $0.368$ & $0.047$ \\
PS1 g	& 	$484.9$	& $0.363$ &	$0.361$	&	$0.109$	&	$0.361$	&	$0.093$	\\%		$0.356$ &	$0.042$ 	& $0.361$ & $0.040$ \\
DES g 	& 	$482.0$	& $0.367$ & 	$0.365$	&	$0.108$	&	$0.365$	&	$0.093$	\\%		$0.359$ &	$0.043$ 	& $0.365$ & $0.042$ \\
LSST r 	& 	$618.5$	& $0.257$ & 	$0.258$	&	$0.089$	&	$0.257$	&	$0.084$	\\%		$0.255$ &	$0.023$ 	& $0.258$ & $0.019$ \\
PS1 r 	& 	$620.1$	& $0.256$ & 	$0.257$	&	$0.089$	&	$0.256$	&	$0.084$	\\%		$0.254$ &	$0.022$ 	& $0.257$ & $0.018$ \\
DES r 	& 	$642.3$	& $0.243$ & 	$0.244$	&	$0.080$	&	$0.243$	&	$0.074$	\\%		$0.241$ &	$0.025$ 	& $0.244$ & $0.020$ \\
VIS	 	& 	$725.0$	& $0.204$ & 	$0.198$	&	$0.081$	&	$0.202$	&	$0.078$	\\%		$0.197$ &	$0.056$	& $0.208$ & $0.050$ \\
LSST i 	& 	$751.8$	& $0.189$ & 	$0.189$	&	$0.065$	&	$0.188$	&	$0.056$	\\%		$0.188$ &	$0.010$ 	& $0.190$ & $0.010$ \\
PS1 i 	&	$753.5$	& $0.188$ & 	$0.188$	&	$0.065$	&	$0.188$	&	$0.056$	\\%		$0.187$ &	$0.010$	& $0.189$ & $0.009$ \\
DES i 	& 	$780.7$	& $0.177$ & 	$0.177$	&	$0.062$	&	$0.177$	&	$0.053$	\\%		$0.176$ &	$0.012$ 	& $0.178$ & $0.011$ \\
LSST z 	& 	$868.3$	& $0.148$ & 	$0.148$	&	$0.053$	&	$0.147$	&	$0.046$	\\%		$0.147$ &	$0.005$ 	& $0.148$ & $0.005$ \\
PS1 z 	& 	$867.4$	& $0.148$ & 	$0.148$	&	$0.053$	&	$0.147$	&	$0.045$	\\%		$0.148$ &	$0.005$ 	& $0.148$ & $0.004$ \\
DES z 	& 	$915.8$	& $0.135$ & 	$0.135$	&	$0.048$	&	$0.134$	&	$0.042$	\\%		$0.134$ &	$0.008$ 	& $0.135$ & $0.007$ \\
LSST y 	&	$966.7$	& $0.123$ & 	$0.123$	&	$0.042$	&	$0.122$	&	$0.037$	\\%		$0.122$ &	$0.004$ 	& $0.123$ & $0.004$ \\
PS1 y 	&	$962.8$	& $0.124$ & 	$0.124$	&	$0.042$	&	$0.123$	&	$0.037$	\\%		$0.123$ &	$0.004$ 	& $0.124$ & $0.004$ \\
DES Y 	&	$986.7$	& $0.118$ & 	$0.119$	&	$0.040$	&	$0.118$	&	$0.036$	\\%		$0.118$ &	$0.005$ 	& $0.119$ & $0.005$ \\
Y 		&	$1055.0$	& $0.106$ & 	$0.105$	&	$0.034$	&	$0.106$	&	$0.031$	\\%		$0.105$ &	$0.012$ 	& $0.107$ & $0.011$ \\
J 		&	$1248.6$	& $0.080$ & 	$0.080$	&	$0.029$	&	$0.080$	&	$0.024$	\\%		$0.079$ &	$0.006$ 	& $0.080$ & $0.006$ \\
H 		&	$1537.0$	& $0.057$ & 	$0.058$	&	$0.020$	&	$0.057$	&	$0.018$	\\%		$0.057$ &	$0.002$ 	& $0.058$ & $0.002$ \\
\hline     
\end{tabular}
\begin{list}{}{}
\small{
\item Galactic extinction expressed in magnitude for line-of-sight $pD = 0.1$.
\item[($1$)] Galactic extinction $A_{\rm X}$ obtained by multiplying $k_{\rm pivot}$ with $pD$
\item[($2$)] Median value of $A_{\rm X}$ for the COSMOS $2604$ models.
\item[($3$)] Dispersion (max$-$min) of $A_{\rm X}$ for the COSMOS $2604$ models
\item[($4$)] Median value of $A_{\rm X}$ for the $37,600$ models fitting CANDELS sources.
\item[($5$)] Dispersion (max$-$min) of $A_{\rm X}$ for the $37,600$ models fitting CANDELS sources.
}
\end{list}
\label{rangeDm}
\end{table*}

\subsection{Impact on photometric corrections}

We probe the impact of the dependence of Galactic absorption with SED by deriving extinctions for a number of SEDs (the COSMOS galaxy model library; Figure~\ref{sed}) which is expanded by dust-reddening and redshifting the templates using a grid of redshift $z = [0 - 2]$ (by bins of $\Delta z = 0.1$) and a grid of intrinsic dust extinction $E_{\rm B-V}^{Int} = [0 - 0.3]$ (by bins of $\Delta E_{\rm B-V}^{Int} = 0.1$) adopting a dust extinction law from \citet{Calzetti2000}. The final template library contains $2604$ templates. 

The Galactic extinction in a filter $X$ ($A_{\rm X} = -2.5$ log$(f_{\rm obs,X}/f_{\rm int,X})$ from equation 1 and 2) is assessed for a set of optical to near-infrared filters relevant for the {\it Euclid} mission and ongoing/upcoming all/half sky surveys: $grizY$ from the Dark Energy Survey (DES) CTIO/DECam\footnote[8]{Total instrumental throughputs including atmosphere are available on the CTIO/DECam website at http://www.ctio.noao.edu/noao/content/dark-energy-camera-decam}, $ugrizy$ from the Large Synoptic Survey Telescope (LSST)\footnote[9]{LSST has two $y$ filters. The filter `$y4$' is adopted here. The models of the commissioned LSST filter total throughputs are retrieved from https://github.com/lsst/throughputs/tree/master/baseline}, $grizY$ from the Panoramic Survey Telescope and Rapid Response System 1 (Panstarrs PS1; Tonry et al.~2012)\nocite{Tonry2012} and three `{\it Euclid}-like' broad-band $YJH$\footnote[10]{We adopt {\it HST}/WFC3 $F105W$, $F125W$ and $F160W$ filter throughputs from http://www.stsci.edu/hst/wfc3/ins\_performance/throughputs as $Y$, $J$ and $H$ filters respectively.}. We also estimate Galactic extinction for a broad hybrid `$riz$' filter mimicking the {\it Euclid} optical VIS throughput planned to cover the wavelength range $\sim550-900$~nm. The modelled VIS throughput was provided by the {\it Euclid} VIS Instrument team.

Figure~\ref{deltam_allps} shows the distribution of Galactic extinction $A_{\rm X}$ for the $2604$ templates, a subset of filters (for readability  purposes) and $pD = 0.1$ along the line of sight. We first derive $A_{\rm X}$ assuming $A_{\rm \lambda} = 0.1 \, k_{\rm \lambda}$, in other words where band-pass corrections and therefore the dependence of $p$ with SED are neglected and only the impact of the SED shape passing through the filter is considered (top panel). We then determine new Galactic extinction estimates once the Milky Way reddening curve has been adequately renormalised using the band-pass correction of a given SED (Figure~\ref{deltam_allps}, bottom). The full impact of SED on the Galactic extinction can be assessed by the dispersion of the distribution. 

Table~\ref{rangeDm} reports the median (column $2$) and dispersion (column $3$) of the Galactic extinction for the $2604$-template library with band-pass correction (i.e.~bottom histograms in Figure~\ref{deltam_allps}). The classically adopted photometric correction obtained by multiplying the reddening along the line of sight (here $0.1$) with $k_{\rm pivot}$ is reported in column $1$ for reference. As already stated earlier in the text, the dependence of Galactic extinction with SED is more significant for the bluer optical wavelengths. For the $u$- and $g$-bands, the extinction can vary by more than $0.1$~mag for regions with $pD > 0.1$, regions that account for $\sim7$\% of the planned {\it Euclid} wide field survey footprint (see section 2). In the $r$-band, variations in $A_{\rm X}$ are greater than $0.08$~mag for $pD = 0.1$.

Up to $pD = 0.3$ (i.e.~the maximum value of reddening we expect along the line of sight of the {\it Euclid} wide field survey; see section 2), $A_{\rm X}$ evolves almost linearly with $pD$ (see Appendix A). $A_{\rm X}$ for other $pD$ values can therefore be linearly derived from Table~\ref{rangeDm}. We note however that $A_{\rm X}$ versus $pD$ strongly deviates from a linear relation ($> 0.01$~mag) from $pD > 0.4$ for the $g$-band, at higher $pD$ values for redder filters.

\subsection{CANDELS}

The goal of this first analysis was to assess the maximum range of extinction potentially encountered by galaxies of different SED; we have thus considered a number of model SED, independently of the probability of finding sources with such SED in the Universe. We have in particular considered  SEDs of passive dusty galaxies ($E_{\rm B-V}^{Int} > 0.2$) that rarely fit observed galaxy colours since early-type galaxies commonly have little dust. We now attempt to provide a more realistic representation of galaxy SED in the Universe. We use the official CANDELS \citep[Cosmic Assembly Near-infrared Deep Extragalactic Legacy Survey;~][]{Grogin2011, Koekemoer2011} multiwavelength photometric catalogues of the GOODS-south \citep{Guo2013} and UDS fields \citep{Galametz2013a}, which respectively contain about $35,000$ sources each. The GOODS-south $17$- and UDS $19$-band photometry was corrected to a first order from Galactic extinction using the reddening value at each source position from the {\it Planck} $E_{\rm B-V}$ map and the value of the F99 reddening curve at the pivot wavelength of each filter. The template-fitting software Phosphoros (Paltani et al.~in prep) was run to derive the photometric redshift and best-fit model of each CANDELS source. 

We use the COSMOS template library, a grid of intrinsic dust extinction $E_{\rm B-V}^{Int} = [0-0.3]$ (with $\Delta E_{\rm B-V}^{Int} = 0.1$ and a Calzetti et al.~2000 dust law). We also apply an extended grid of redshift $0 < z < 6$ (with $\Delta z = 0.01$) to exclude $z > 2$ sources and the prescription of \citet{Meiksin2006} to take into account the photometric extinction of sources by the intergalactic medium absorption (IGM). We only consider sources with photometric redshift $z_{\rm phot} < 2$ and sources with flag $= 0$ \citep[][Appendix B]{Galametz2013a} and $z_{\rm phot} > 0.02$ to avoid potential stars and artefacts. The final GOODS-south $+$ UDS sample contains $37,600$ sources. The Galactic extinction is derived for each best-fit model SED, $pD = 0.1$ and for the same set of filters as in section 3.5 (optical LSST, DES, PS1 and {\it Euclid}-like filters). Table~\ref{rangeDm} (columns $4$ and $5$) reports the median and dispersion of $A_{\rm X}$ for the $37,600$ SEDs. The range of extinction is only mildly  less extended for this more representative set of models. The Galactic extinction in $u$- and $g$-bands can vary to more than $0.09$~mag for $pD = 0.1$ and up to $0.08$ in the $r$-band.

\begin{figure}
\begin{center}
\includegraphics[width = 11.5cm,bb= 60 30 650 800]{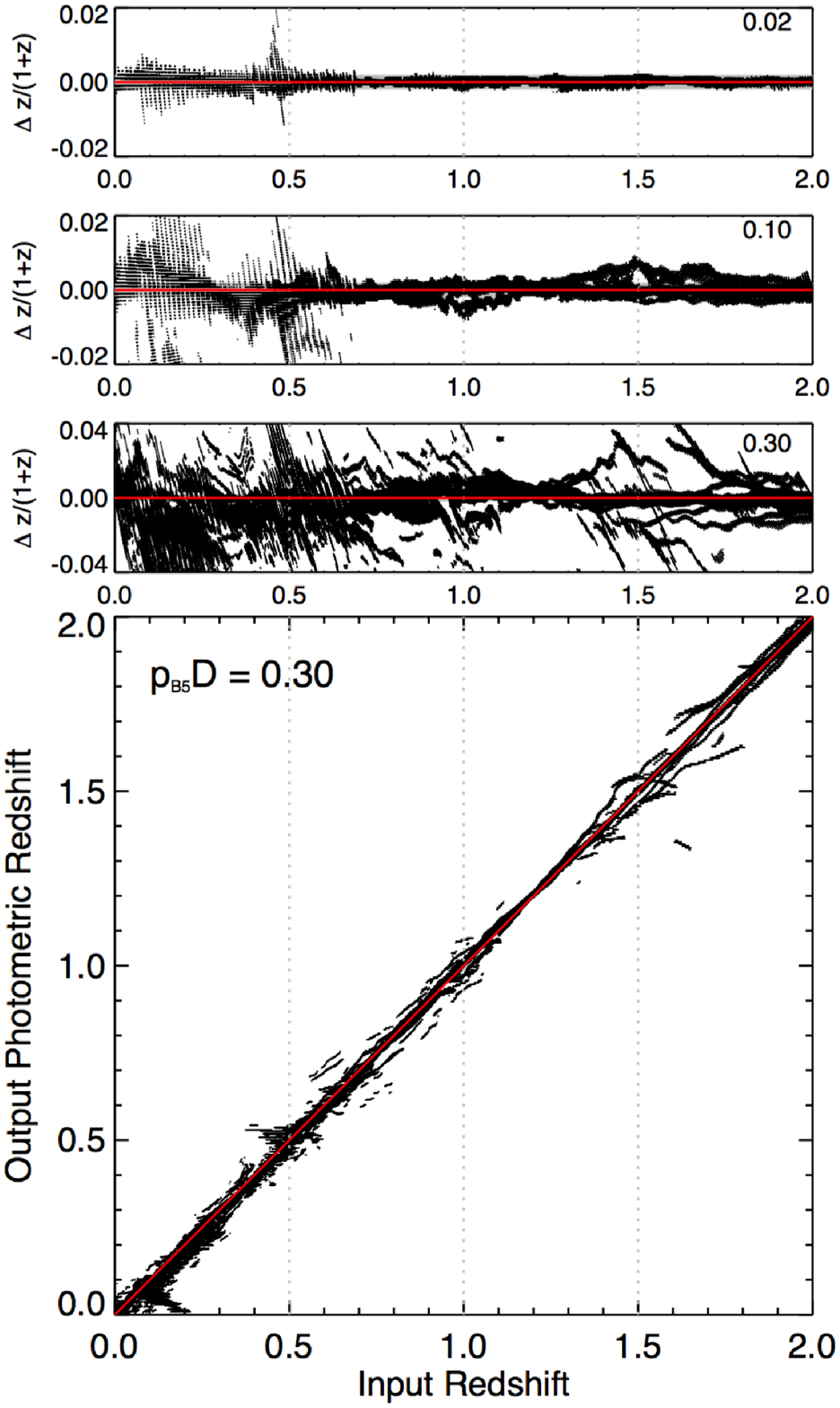}
\end{center}
\caption{Differences between input redshift of the model catalogues and output photometric redshift estimates. The top three panels show the differences $\Delta z$ normalised by $(1+z)$ as a function of input redshift for three investigated $p_{\rm B5}D$ extinction values: $p_{\rm B5}D = 0.02$, $0.1$ and $0.3$ ($pD$ values indicated in the upper right corner). The grey zone indicates the constrains on the photo-z bias required to achieve the precisions on the cosmological parameters set by the upcoming weak-lensing surveys such as {\it Euclid} i.e.~$\langle z \rangle < 0.002(1+z)$. The bottom panel shows the input versus output redshifts for $p_{\rm B5}D = 0.3$. Vertical grey lines at $z = 0.5$, $1$ and $1.5$ are plotted to guide the eye. The impact of neglecting the Galactic absorption dependence with SED can directly be assessed from the divergences from the identity line.}
\label{zp}
\end{figure}

\section{Impact on photometric redshift determination}

An inaccurate (under/overestimated) correction of Galactic extinction will severely affect photometric measurements and thus bias the determination of photometric redshifts. Since the inaccuracies are SED-dependent, the biases are expected to be more critical for some source types and/or for some redshifts than others, for example when some of the SED characteristic breaks pass in certain filters. They will therefore lead to systematic uncertainties in the estimation of cosmological parameters when conducting weak lensing analysis. 

In this section, we evaluate the errors on photometric redshifts that ensue from neglecting the dependence of Galactic absorption with SED. We build mock flux catalogues of sources with known intrinsic SED using the COSMOS library, a grid of intrinsic dust extinction $E_{\rm B-V}^{Int} = [0-0.3]$ (with $\Delta E_{\rm B-V}^{Int} = 0.1$ and a Calzetti et al.~2000 dust law) and a grid of redshift $z = [0-2]$. The effects mention in the following can be of the order of a few $0.1\%(1+z)$ so the grid of redshift was refined to $\Delta z = 0.001$. 

We construct three catalogues for three $p_{\rm B5}D$ values along the line of sight: $0.02$ (classical extinction in the direction of the most commonly studied extragalactic surveys), $0.1$ (high extinction) and $0.3$ (the expected upper limit of reddening in the {\it Euclid} wide field survey). In other words, we assume that these value of extinction mimics the pixel value read at the position of a source on a reddening map calibrated using B5 stars. Since the source SED is known, we can appropriately apply Galactic absorption by multiplying the template by $10^{-0.4 pD k_{\rm \lambda}}$ (see equation 2) where $p_{\rm sed}D = p_{\rm B5}D / {\rm bpc}_{\rm sed}$, meaning correctly corrected by the appropriate band-pass correction. We then derive for each template the fluxes within a set of filters mimicking the bands planned (not yet guaranteed however) in the upcoming {\it Euclid} wide field and complementary optical ground-based surveys ($ugrizy$). We adopt for this analysis the filter throughputs of the DES survey and $YJH$ following footnote 10. We assume an infinite signal-to-noise on the photometry.

All fluxes in a given filter are then systematically corrected by the commonly adopted estimate of Galactic extinction $pD \, k_{\rm pivot}$ where $pD = 0.02$, $0.1$ and $0.3$ and $k_{\rm pivot}$ the value of the absorption law at the pivot wavelength of each filter. We subsequently attempt to recover from this `classic extinction-corrected' photometry the input redshift of the models using the template-fitting software Phosphoros and the same library of templates (same template, intrinsic dust and redshift grid, used to generate the catalogues). 

Figure~\ref{zp} shows the differences between the initial redshift of the templates and the output photometric redshifts provided by Phosphoros for the different values of extinction along the line-of-sight. Large deviations from the identity line are observed especially at $z < 0.7$. In that redshift range and for the respective lines of sight of $pD = 0.02$, $0.1$ and $0.3$, $\sim15$, $60$ and $85$\% of the sources have a photometric redshift estimate discrepant of $> 0.2\%(1+z)$ from the initial template redshift. As mentioned earlier, the scientific goals of upcoming weak-lensing surveys require an accuracy on the photometric redshift bias below this threshold of $0.2\%(1+z)$. For $pD = 0.1$, the bias derived in the redshift bins $z = [0.3-0.4]$ and $z = [0.4-0.5]$ is $\langle z \rangle / (1+z) \sim 0.3\%$ and $\sim 0.2\%$ respectively. The discrepancies are more moderate beyond $z > 0.7$. We note however that, for $pD = 0.1$, an average of $20$\% of the template redshift are not recovered to an accuracy of $0.002 (1+z)$. As a sanity check, we reiterate the photometric redshift estimation using the template-fitting code EAZY \citep{Brammer2008} and find consistent results.

\begin{figure}
\begin{center}
\includegraphics[width = 9cm,bb = 20 500 820 720]{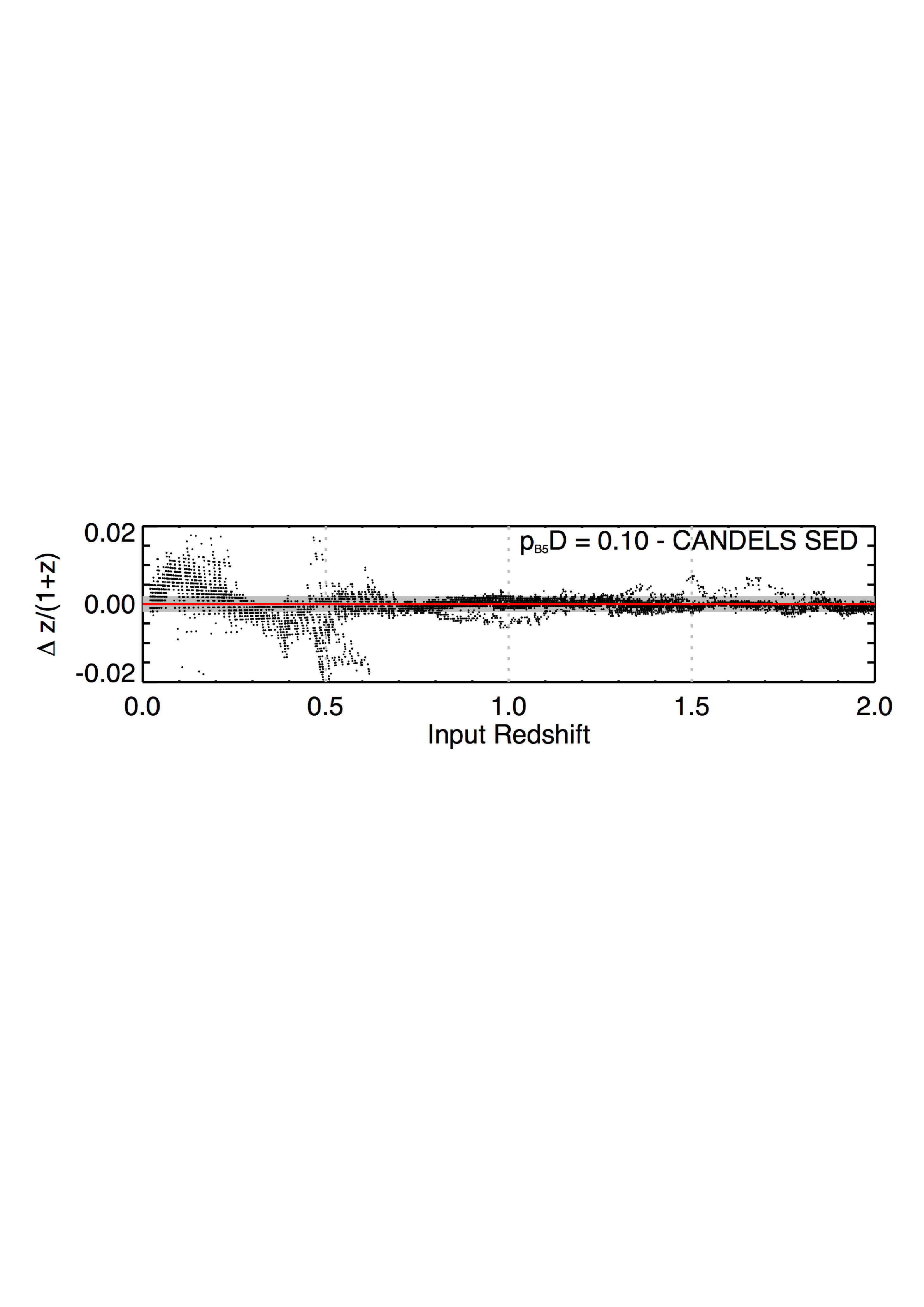}
\end{center}
\caption{Same as Figure~\ref{zp} (second panel, $p_{\rm B5}D = 0.1$) but for a more realistic representation of galaxy SEDs in the Universe i.e.~weighted by the probability of finding such SED in the CANDELS GOODS-south and UDS fields.}
\label{zpcandels}
\end{figure}

We attempt to derive a more realistic estimate of the impact of Galactic extinction inaccuracies on photometric redshifts by weighting the template library by a more representative distribution of SED. As in section 3.6, we make use of the CANDELS multi-wavelength catalogues to derive the best fit SED of sources in the GOODS-south and UDS fields. We fit the CANDELS sources using the same template library, intrinsic dust and redshift binning as the input catalogue over the redshift range $z = [0-6]$. As Ly$\alpha$ enters the $u$-band at $z \ge 2$, we also adopt the IGM prescription of \citet{Meiksin2006} to take into consideration the extinction of source photometry due to intergalactic medium absorption. We then only consider the sources with $0.02 < z_{\rm phot} < 2$ and flag $= 0$ (see section 3.6). Figure~\ref{zpcandels} shows the differences between the redshift of the templates and the output photometric redshifts provided by Phosphoros for $pD = 0.1$ (similar to Figure~\ref{zp}, second panel) but weighted by the probability of finding a source with such SED in the CANDELS fields. We find similar results as for the whole template library. The bias on photometric redshifts is even strengthened for $0.3 < z < 0.5$ with $\Delta z / (1+z) < -3\%$. These results demonstrate that not taking into account the dependence of SED for Galactic extinction estimates could lead to substantial inaccuracies in the estimation of photometric redshifts.

\section{Conclusions and perspectives}

We evaluate the impact of a source SED on Galactic extinction. The use of calibration sources of a specific SED with broad-band filters systematically bias the calibration of the Galactic absorption law and band-pass corrections are required to adequately renormalise the absorption law for a given SED. We assess the range of band-pass corrections that could be expected from a library of extragalactic sources derived from the {\rm LePhare} COSMOS template library and a grid of intrinsic dust and redshift. We use the Fitzpatrick (1999) Milky Way absorption law that was calibrated using the colour excesses of B5 stars and find that band-pass corrections required to renormalise a B5 star-calibrated law can be in the range $0.8 < {\rm bpc}_{\rm sed} < 1.2$; in other words, the product of the absorption law by the reddening along the line-of-sight (i.e.~the `$E_{\rm B-V}$' read on a reddening map) needs to be scaled up/down by up to $20$\% to take into consideration the SED of a source. 

One of the direct consequences of neglecting these band-pass corrections is that the scaling of a reddening map obtained from a dust map using the colour excesses of sources of a given SED will be SED-dependent. We illustrate this point by deriving the dust-to-reddening scaling factors for three different types of sources i.e.~three distinct SED. For this work, we use $z < 0.4$ luminous red galaxies from SDSS divided in three redshift bins ($[0.15-0.25]$, $[0.25-0.3]$ and $[0.3-0.35]$). The value of Galactic extinction along the line-of-sight of the sources is derived by matching their SDSS photometry with a reference LRG template reddened by a grid of extinction. We then linearly correlate the values of the radiance $R$ map produced by the {\it Planck} team (P14) in the direction of the sources with the estimate of reddening. This analysis is repeated for each studied LRG redshift bin. The differences in scaling factors for the three samples of LRG can be predicted by the band-pass corrections derived directly from the SED. We also recover the dust-to-reddening scaling factor that was derived by the {\it Planck} team (P14) using quasars at $0.7 < z < 1.7$ to convert their radiance $R$ map into the released `$E_{\rm B-V}$' reddening map. 

We then estimate the range of Galactic extinction corrections to be expected for an SED library of extragalactic sources within a number of optical-to-near infrared filters pertinent to on-going or upcoming cosmological surveys. We find that Galactic extinction corrections can vary by up to $0.1$~mag in the optical bands for lines of sight of medium-to-high extinction $pD > 0.1$ which represents about $\sim8$\% of the {\it Euclid} Wide survey planned footprint. Corrections can vary up to $0.02$~mag for lines of sight $pD = 0.02$, the extinction average value commonly encountered in the direction of the most well studied extragalactic surveys. The impact of a source SED for Galactic extinction in the near infrared wavelengths is less significant but still found up to $0.04$~mag in the $Y$-band for $pD = 0.1$. The introduction of biases in the estimation of Galactic extinction photometric corrections naturally result in biases in the estimates of source photometric redshifts that could be dramatic for future cosmological surveys, especially in high extinction lines of sight.

Appendix A presents an example of prescription of the Galactic extinction in template-fitting codes used for the determination of photometric redshifts. It describes in particular the Galactic extinction recipe that will be implemented in the template-fitting code of the {\it Euclid} Mission, Phosphoros. Upcoming works will focus on illustrating the improvement on photometric redshifts induced by the use of this exact prescription of Galactic extinction. They will particularly concentrate on testing photometric redshift estimates for sources along lines of sight of high extinction, classically avoided by extragalactic wide-field surveys. Ultimately, we will be releasing a set of online routines and applets that will provide for a given line of sight and SED the estimate of Galactic extinction for a set of common filters\footnote[11]{In the meantime, please feel free to contact the authors for early access to the routines.}. 

Appendix B investigates the impact of uncertainties of the extinction law and in particular of its characteristic total-to-selective ratio $R_{\rm V}$ on Galactic extinction estimates. We show that although the uncertainties on $R_{\rm V}$ do not cause systematic biases, they are reducing the precision of photometric measurements and by extension of photometric redshifts. Appendix C confronts our analysis to SFD98 and raises again the impact of the adopted extinction law prescription for Galactic extinction estimates.

In the context of template-fitting codes, the a-priori knowledge of the template SED used to model the observed photometry allows to properly estimate Galactic absorption. Such implementation is however not easily applicable for non-template-based photometric redshift recipes such as machine-learning architectures. Machine-learning algorithms learn how to combine galaxy photometric measurements \citep[and other galaxy features; see e.g.~][]{Hoyle2015} to estimate source distances using training samples of sources with known redshift. Unless such machine is trained with spectroscopically confirmed sources that cover a wide and `complete' range of Galactic reddening, it will not be able to learn how to take into account the impact of Galactic extinction on photo-z estimates correctly. A possible alternative could be to combine the strength of both template-fitting codes and machine-learning algorithms and use the first guess of the source type or SED from template-fitting to correct (to a first order at least) the source photometry from Galactic extinction. Complementarily, Masters et al.~(2015)\nocite{Masters2015} recently studied the colour-redshift distribution of galaxies based on a self-organising map (SOM) analysis with each cell of the SOM corresponding approximately to a characteristic source SED. One of their main science drivers was to investigate the regions of their SOM lacking spectroscopic redshifts and estimate the missing spectroscopy potentially required to calibrate photometric redshifts. The analysis was conducted in the COSMOS field and therefore with low Galactic extinction along the line-of-sight. It would be interesting to investigate how the Galactic extinction and its dependence with source SED would affect the colour-redshift distribution.

Finally, the success of upcoming weak-lensing analyses does not only depend on the estimation of photometric redshifts of high precision. It also requires an accurate determination of the shape measurement of sources, more specifically a precise measure of their distortion, the so-called shear signal. In the light of the dependence of Galactic extinction with galaxy SED/colour, we also raise concern about the fact that the amount of Galactic extinction will also be sensitive to the colour gradients within the source itself and that it might bias the determination of source shape measurements.

\begin{acknowledgements}
Audrey Galametz acknowledges support by the Deutsches Zentrum f{\"u}r Luft- und Raumfahrt (DLR) grant 50 QE 1101. We are grateful to Natascha Greisel and Stella Seitz for having provided the luminous red galaxy template models used in section 3.4 and to Salvador Salazar Albornoz for his help handling the SDSS-III/BOSS database. We also would like to thank the ESA staff and Euclid Consortium members for useful discussions about this work and many suggestions that have undoubtedly helped improve the paper.
\end{acknowledgements}

\bibliographystyle{apj}
\bibliography{biblio}

\begin{thebibliography}{44}
\expandafter\ifx\csname natexlab\endcsname\relax\def\natexlab#1{#1}\fi

\bibitem[{{Berry} {et~al.}(2012){Berry}, {Ivezi{\'c}}, {Sesar},
  {et~al.}}]{Berry2012}
{Berry}, M., {Ivezi{\'c}}, {\v Z}., {Sesar}, B., {et~al.} 2012, \apj, 757, 166

\bibitem[{{Bordoloi} {et~al.}(2010){Bordoloi}, {Lilly}, \&
  {Amara}}]{Bordoloi2010}
{Bordoloi}, R., {Lilly}, S.~J., \& {Amara}, A. 2010, \mnras, 406, 881

\bibitem[{{Brammer} {et~al.}(2008){Brammer}, {van Dokkum}, \&
  {Coppi}}]{Brammer2008}
{Brammer}, G.~B., {van Dokkum}, P.~G., \& {Coppi}, P. 2008, \apj, 686, 1503

\bibitem[{{Bruzual} \& {Charlot}(2003)}]{Bruzual2003}
{Bruzual}, G. \& {Charlot}, S. 2003, \mnras, 344, 1000

\bibitem[{{Burstein} \& {Heiles}(1978)}]{Burstein1978}
{Burstein}, D. \& {Heiles}, C. 1978, \apj, 225, 40

\bibitem[{{Calzetti} {et~al.}(2000){Calzetti}, {Armus}, {Bohlin},
  {et~al.}}]{Calzetti2000}
{Calzetti}, D., {Armus}, L., {Bohlin}, R.~C., {et~al.} 2000, \apj, 533, 682

\bibitem[{{Cardelli} {et~al.}(1989){Cardelli}, {Clayton}, \&
  {Mathis}}]{Cardelli1989}
{Cardelli}, J.~A., {Clayton}, G.~C., \& {Mathis}, J.~S. 1989, \apj, 345, 245

\bibitem[{{Dahlen} {et~al.}(2013){Dahlen}, {Mobasher}, {Faber},
  {et~al.}}]{Dahlen2013}
{Dahlen}, T., {Mobasher}, B., {Faber}, S.~M., {et~al.} 2013, \apj, 775, 93

\bibitem[{{Dark Energy Survey Collaboration}(2016)}]{DES2016}
{Dark Energy Survey Collaboration}. 2016, \mnras, 460, 1270

\bibitem[{{de Jong} {et~al.}(2013){de Jong}, {Kuijken}, {Applegate},
  {et~al.}}]{deJong2013}
{de Jong}, J.~T.~A., {Kuijken}, K., {Applegate}, D., {et~al.} 2013, Messenger,
  154, 44

\bibitem[{{Fitzpatrick}(1999)}]{Fitzpatrick1999}
{Fitzpatrick}, E.~L. 1999, \pasp, 111, 63

\bibitem[{{Galametz} {et~al.}(2013){Galametz}, {Grazian}, {Fontana},
  {et~al.}}]{Galametz2013a}
{Galametz}, A., {Grazian}, A., {Fontana}, A., {et~al.} 2013, \apjs, 206, 10

\bibitem[{{Greisel} {et~al.}(2013){Greisel}, {Seitz}, {Drory},
  {et~al.}}]{Greisel2013}
{Greisel}, N., {Seitz}, S., {Drory}, N., {et~al.} 2013, \apj, 768, 117

\bibitem[{{Grogin} {et~al.}(2011){Grogin}, {Kocevski}, {Faber},
  {et~al.}}]{Grogin2011}
{Grogin}, N.~A., {Kocevski}, D.~D., {Faber}, S.~M., {et~al.} 2011, \apjs, 197,
  35

\bibitem[{{Guo} {et~al.}(2013){Guo}, {Ferguson}, {Giavalisco},
  {et~al.}}]{Guo2013}
{Guo}, Y., {Ferguson}, H.~C., {Giavalisco}, M., {et~al.} 2013, \apjs, 207, 24

\bibitem[{{Hearin} {et~al.}(2010){Hearin}, {Zentner}, {Ma}, \&
  {Huterer}}]{Hearin2010}
{Hearin}, A.~P., {Zentner}, A.~R., {Ma}, Z., \& {Huterer}, D. 2010, \apj, 720,
  1351

\bibitem[{{Hildebrandt} {et~al.}(2010){Hildebrandt}, {Arnouts}, {Capak},
  {et~al.}}]{Hildebrandt2010}
{Hildebrandt}, H., {Arnouts}, S., {Capak}, P., {et~al.} 2010, \aap, 523, A31

\bibitem[{{Hoekstra} {et~al.}(2002){Hoekstra}, {Yee}, {Gladders},
  {et~al.}}]{Hoekstra2002}
{Hoekstra}, H., {Yee}, H.~K.~C., {Gladders}, M.~D., {et~al.} 2002, \apj, 572,
  55

\bibitem[{{Hoyle} {et~al.}(2015){Hoyle}, {Rau}, {Zitlau}, {et~al.}}]{Hoyle2015}
{Hoyle}, B., {Rau}, M.~M., {Zitlau}, R., {et~al.} 2015, \mnras, 449, 1275

\bibitem[{{Hu}(1999)}]{Hu1999}
{Hu}, W. 1999, \apjl, 522, L21

\bibitem[{{Ilbert} {et~al.}(2009){Ilbert}, {Capak}, {Salvato},
  {et~al.}}]{Ilbert2009}
{Ilbert}, O., {Capak}, P., {Salvato}, M., {et~al.} 2009, \apj, 690, 1236

\bibitem[{{Ivezic} {et~al.}(2008){Ivezic}, {Tyson}, {et~al.}}]{Ivezic2008}
{Ivezic}, Z., {Tyson}, J.~A., {et~al.} 2008, ArXiv:0805.2366

\bibitem[{{Koekemoer} {et~al.}(2011){Koekemoer}, {Faber}, {Ferguson},
  {et~al.}}]{Koekemoer2011}
{Koekemoer}, A.~M., {Faber}, S.~M., {Ferguson}, H.~C., {et~al.} 2011, \apjs,
  197, 36

\bibitem[{{Kriek} \& {Conroy}(2013)}]{Kriek2013}
{Kriek}, M. \& {Conroy}, C. 2013, \apjl, 775, L16

\bibitem[{{Laureijs} {et~al.}(2011){Laureijs}, {Amiaux}, {Arduini},
  {et~al.}}]{Laureijs2011}
{Laureijs}, R., {Amiaux}, J., {Arduini}, {et~al.} 2011, ArXiv:1110.3193

\bibitem[{{Ma} {et~al.}(2006){Ma}, {Hu}, \& {Huterer}}]{Ma2006}
{Ma}, Z., {Hu}, W., \& {Huterer}, D. 2006, \apj, 636, 21

\bibitem[{{Masters} {et~al.}(2015){Masters}, {Capak}, {Stern},
  {et~al.}}]{Masters2015}
{Masters}, D., {Capak}, P., {Stern}, D., {et~al.} 2015, \apj, 813, 53

\bibitem[{{Meiksin}(2006)}]{Meiksin2006}
{Meiksin}, A. 2006, \mnras, 365, 807

\bibitem[{{M{\"o}rtsell}(2013)}]{Mortsell2013}
{M{\"o}rtsell}, E. 2013, \aap, 550, A80

\bibitem[{{Noll} {et~al.}(2009){Noll}, {Burgarella}, {Giovannoli},
  {et~al.}}]{Noll2009}
{Noll}, S., {Burgarella}, D., {Giovannoli}, E., {et~al.} 2009, \aap, 507, 1793

\bibitem[{{O'Donnell}(1994)}]{O'Donnell1994}
{O'Donnell}, J.~E. 1994, \apj, 422, 158

\bibitem[{{Padmanabhan} {et~al.}(2005){Padmanabhan}, {Budav{\'a}ri},
  {Schlegel}, {et~al.}}]{Padmanabhan2005}
{Padmanabhan}, N., {Budav{\'a}ri}, T., {Schlegel}, D.~J., {et~al.} 2005,
  \mnras, 359, 237

\bibitem[{{Padmanabhan} {et~al.}(2008){Padmanabhan}, {Schlegel}, {Finkbeiner},
  {et~al.}}]{Padmanabhan2008}
{Padmanabhan}, N., {Schlegel}, D.~J., {Finkbeiner}, D.~P., {et~al.} 2008, \apj,
  674, 1217

\bibitem[{{Peek} \& {Graves}(2010)}]{Peek2010}
{Peek}, J.~E.~G. \& {Graves}, G.~J. 2010, \apj, 719, 415

\bibitem[{{Planck Collaboration}(2014)}]{Planck2014}
{Planck Collaboration}. 2014, \aap, 571, A11

\bibitem[{{Schlafly} \& {Finkbeiner}(2011)}]{Schlafly2011}
{Schlafly}, E.~F. \& {Finkbeiner}, D.~P. 2011, \apj, 737, 103

\bibitem[{{Schlafly} {et~al.}(2010){Schlafly}, {Finkbeiner}, {Schlegel},
  {et~al.}}]{Schlafly2010}
{Schlafly}, E.~F., {Finkbeiner}, D.~P., {Schlegel}, D.~J., {et~al.} 2010, \apj,
  725, 1175

\bibitem[{{Schlafly} {et~al.}(2014){Schlafly}, {Green}, {Finkbeiner},
  {et~al.}}]{Schlafly2014}
{Schlafly}, E.~F., {Green}, G., {Finkbeiner}, D.~P., {et~al.} 2014, \apj, 789,
  15

\bibitem[{{Schlegel} {et~al.}(1998){Schlegel}, {Finkbeiner}, \&
  {Davis}}]{Schlegel1998}
{Schlegel}, D.~J., {Finkbeiner}, D.~P., \& {Davis}, M. 1998, \apj, 500, 525

\bibitem[{{Schneider} {et~al.}(2010){Schneider}, {Richards}, {Hall},
  {et~al.}}]{Schneider2010}
{Schneider}, D.~P., {Richards}, G.~T., {Hall}, P.~B., {et~al.} 2010, VizieR
  Online Data Catalog, 7260, 0

\bibitem[{{Tokunaga} \& {Vacca}(2005)}]{Tokunaga2005}
{Tokunaga}, A.~T. \& {Vacca}, W.~D. 2005, \pasp, 117, 421

\bibitem[{{Tonry} {et~al.}(2012){Tonry}, {Stubbs}, {Lykke},
  {et~al.}}]{Tonry2012}
{Tonry}, J.~L., {Stubbs}, C.~W., {Lykke}, K.~R., {et~al.} 2012, \apj, 750, 99

\bibitem[{{Vanden Berk} {et~al.}(2001){Vanden Berk}, {Richards}, {Bauer},
  {et~al.}}]{VandenBerk2001}
{Vanden Berk}, D.~E., {Richards}, G.~T., {Bauer}, A., {et~al.} 2001, \aj, 122,
  549

\bibitem[{{York} \& {The SDSS Collaboration}(2000)}]{York2000}
{York}, D.~G. \& {The SDSS Collaboration}. 2000, \aj, 120, 1579

\end{thebibliography}

\clearpage

\begin{appendix}

\section{Prescription of Galactic extinction for template-fitting codes}

In photometric redshift template-fitting codes, the best redshift estimate is derived by comparing the observed photometry of a source with the one computed from a library of templates. Classically, the photometric measurements are first corrected from Galactic extinction using the $E_{\rm B-V}$ along the line of sight multiplied by the value of the Milky Way absorption law at the filter pivot wavelength and then confronted with the template fluxes. The dependence of Galactic extinction with SED is therefore traditionally neglected, mostly due to the lack of a-priori knowledge on a source SED. 

This approximation can however be avoided in the context of template-fitting codes by applying the Galactic extinction to the template flux catalogue instead of the observed photometry catalogue. The knowledge of each SED in a template-fitting code allows to adequately quantify the extinction undergone by a source with such SED. We describe in the following the prescription of Galactic extinction that will be adopted within the Phosphoros photometric redshift code for reference and as an example for a proposed improvement of existing template-fitting algorithms. 

\begin{figure}
\begin{center}
\includegraphics[width = 9cm,bb = 20 30 550 470]{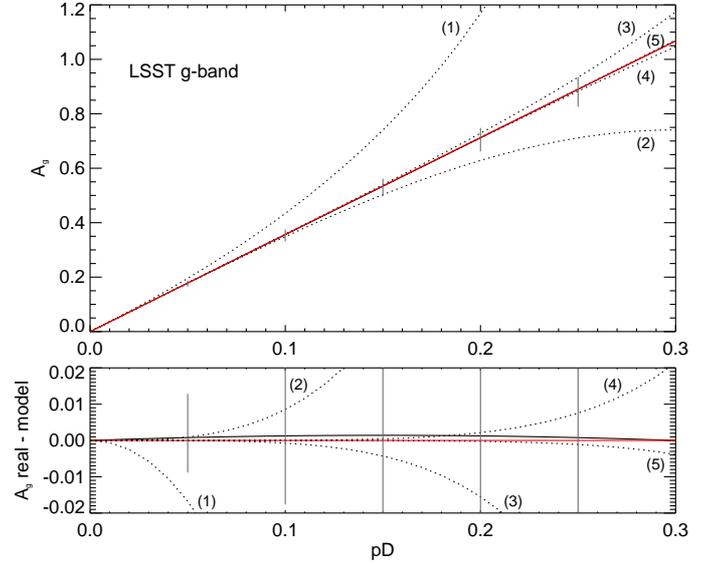}
\end{center}
\caption{{\it Top panel:} Galactic extinction in the LSST $g$ filter (in magnitude) as a function of the value of $pD$ along the line of sight for an elliptical galaxy with no intrinsic dust ($E_{\rm B-V}^{Int} = 0$) at $z = 0.1$ (in red). A linear approximation $A_g = a \, pD$ where $a$ the slope between $(0,0)$ and $(0.3, A_g(0.3))$ is shown in black. The first cumulative orders of the mathematical decomposition of this function in Taylor series are shown by the dotted lines (see equation A.2; $(1) = \sum\limits_{\rm n=0}^{1}$, $(2) = \sum\limits_{\rm n=0}^{2}$ etc.). The grey vertical lines indicate the range of Galactic extinction at fixed $pD$ for a library of $2604$ extragalactic source SED (see section 3.5). {\it Bottom:} Differences between the real value of Galactic extinction computed from equations 1 and 2 and the linear and Taylor series approximations. The scatter of extinction values for a range of SEDs is also reproduced in grey.}
\label{phos}
\end{figure}

One of the major components of template-fitting codes is the build-up of the template flux catalogue where a library of restframe SEDs is respectively dust-reddened (assuming a range of intrinsic dust values and absorption laws), redshifted and finally integrated through a set of filters. For an individual galaxy with a given line of sight extinction, one can apply the Galactic extinction directly on the SED before integration through the filters (see equation 2). The Galactic extinction is however inhomogeneous along the line of sight of large sky surveys and it would be far too time- and computer-demanding to re-derive the flux template library for all values of reddening. The Galactic extinction is therefore parametrised for each SED and each filter as a function of $pD$. Figure.~\ref{phos} presents the extinction in the LSST $g$-band $A_g$ ($= -2.5log(f_{\rm obs,g}) + 2.5log(f_{\rm int,g})$) sustained by an elliptical galaxy at $z = 0.1$ as a function of $pD$. We attempt to model the value of $A_X$ as a function of the sole parameter $pD$. In equation 2, 

\begin{equation}
10^{-0.4A_\lambda} = 10^{-0.4 pD  k_\lambda} = exp [-0.4 ln(10) pD k_\lambda] = \sum\limits_{\rm n=0}^{\infty} \frac{t^n}{n!} (pD)^n \end{equation}

where $t = -0.4ln(10)k_{\rm \lambda}$. And therefore the Galactic extinction in a filter $X$ is given by  

\begin{equation}
A_X = -2.5log\left( \sum\limits_{\rm n=0}^{\infty} (pD)^n \frac{\int_{\rm X} \frac{t^n}{n!} f_{\rm sed}(\lambda)F_X(\lambda) d\lambda}{\int_{\rm X} F_X(\lambda) \frac{c}{\lambda^2} d\lambda} \right) + 2.5log(f_{\rm int,X}).
\end{equation}

The first cumulative ($\sum\limits_{\rm n=0}^{1}$, $\sum\limits_{\rm n=0}^{2}$ etc.) orders of the decomposition of this function in Taylor series are shown in Figure~\ref{phos}. We observe that a correct approximation of the Galactic extinction is not reached for $pD < 0.3$ before the $5^{th}$ (cumulative) order of the decomposition. The implementation of such approximation would therefore require the storage of at least $5-6$ numbers to characterise extinction for each SED and each filter which could be extremely memory-demanding. 

We therefore adopt a simpler representation, a linear approximation of the Galactic extinction, in the filter $X$, $A_X = a_{\rm sed\_X} \, pD$ of the function between $(pD,A_X(pD)) = (0,0)$ and $(0.3,A_X(0.3))$. We adopt $pD = 0.3$ as it is the upper limit of reddening expected within the {\it Euclid} wide field survey footprint. Both approximations are plotted in Figure~\ref{phos} which also shows the differences between the real value of Galactic extinction and the linear and Taylor series approximations. In this case example, the linear approximation recovers the Galactic extinction to $0.001$~mag. The range of Galactic extinction expected for fixed values of $pD$ ($= 0.05$, $0.1$ etc.) is also plotted for a variety of SED (see section 3.5; $2604$ SEDs) for comparison. We mention that at $pD = 0.1$, the plotted range corresponds to the dispersion of the histogram for the LSST $g$ filter in Figure~\ref{deltam_allps} (top panel i.e.~not taking into account the additional scatter introduced by correcting the reddening value by the band-pass correction). We note that the error of a linear approximation is orders of magnitude smaller than the scatter introduced by the dependence of extinction with the SED.

While building the template catalogue, the template-fitting code therefore needs to derive the band-pass correction corresponding to each SED (bpc$_{\rm sed}$; see 3.2) and the linear coefficient `$a_{\rm sed\_filter}$' for each SED and each filter that will allow to derive the Galactic extinction correction for a given $pD$. The value of reddening read on the P14 map at the source position is divided by bpc$_{\rm sed}$ and  multiplied by bpc$_{\rm P14} = 1.018$ --- since the P14 map is calibrated with quasars --- to obtain the actual $pD$ value for a given SED. The template photometry is then corrected in each filter using $A_X$ ($= a_{\rm sed\_X} \, pD$).

\section{Impact of the uncertainties of the total-to-selective ratio $R_{\rm V}$}

Extinction depends on the properties (size and composition) of dust particles along the line of sight which implies that the extinction curve derived for different directions in the Milky Way is line-of-sight dependent. Numerous studies have however shown that, along most lines of sight, the extinction law can be parametrised in terms of a unique value, the total-to-selective extinction ratio $R_{\rm V} = A_{\rm V} / E_{\rm B-V}$ (Cardelli, Clayton \& Mathis 1989). In brief, $R_{\rm V}$ characterises the composition of dust along the line of sight while $pD$ (or $E_{\rm B-V}$) quantifies the amount of dust in the Milky Way along the observed direction. $R_{\rm V}$ has been found to vary from $2.2$ to up to $5.8$ for dense molecular clouds with a mean value of $R_{\rm V} = 3.1$ for a diffuse interstellar medium. Few works have attempted to derive the Milky Way extinction law along high galactic latitudes, relevant to extragalactic studies. M{\"o}rtsell~(2013) however recently used cosmological sources (quasars, BCG and LRG) to estimate $R_{\rm V}$ in high latitude regions and found a consistent $R_{\rm V} \sim 3.1-3.2$. Their analysis did not reveal any major spatial variations in the dust properties across their probed lines of sight. Their results echoed the findings of \citet{Berry2012} who also recovered a median $R_{\rm V} \sim 3.1$ using the SDSS-2MASS photometry of stars at high Galactic latitudes. They derived a precision of $0.2$ for the value of $R_{\rm V}$.

To date, there has been no systematic mapping of the properties of dust grains in all directions of the Milky Way due to the source density that such studies would entail. Although it may have been natural to hypothesise that, to some extent, $R_{\rm V}$ could correlate with the dust column density along the line of sight, M{\"o}rtsell~(2013) did not find any evidence for such correlation in their sample of quasars when segregated in two regions of low and high dust column density. It is therefore unfeasible at this stage to predict, for a given line of sight, an exact estimate of the Milky Way extinction. To take into account the statistical uncertainty in $R_{\rm V}$ from past calibration works (i.e.~$= 3.1 \pm 0.2$), one therefore needs to add an error to the Galactic extinction estimate derived for a fixed value of $R_{\rm V}$ (e.g.~for $R_{\rm V} = 3.1$ in this work). This appendix explores the impact of the uncertainties of $R_{\rm V}$ on Galactic extinction predictions and photo-z estimates.

\begin{figure}
\begin{center}
\includegraphics[width = 9cm,bb = 0 45 470 420]{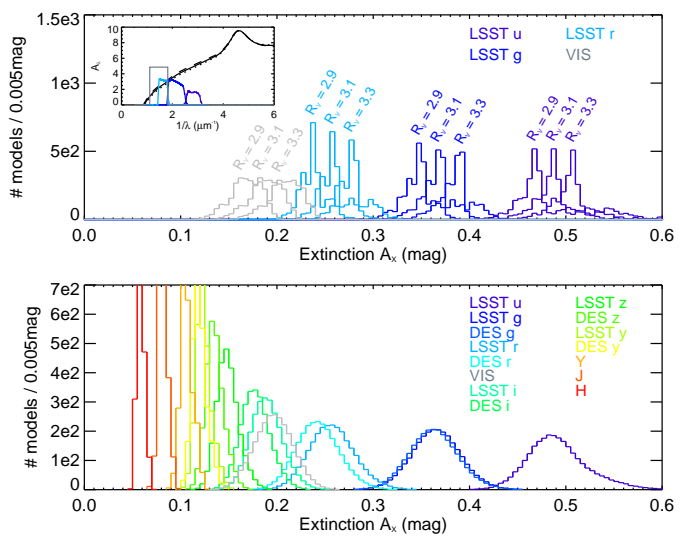}
\end{center}
\caption{Distribution of Galactic extinction $A_{\rm X}$ (in magnitude) for $p_{\rm B5}D = 0.1$ for a library of $2604$ extragalactic source SEDs and a set of optical-to-near infrared filters. {\it Top panel:} Same as Figure~\ref{deltam_allps} for three values of $R_{\rm V}$ ($= 2.9$, $3.1$ and $3.3$). The inset panel shows the F99 extinction curves for $R_{\rm V} = 2.9$ (dotted), $3.1$ (solid) and $3.3$ (dashed) along with the filter throughputs. {\it Bottom panel:} Median of $100$ realisations of the distribution of Galactic extinction $A_{\rm X}$ for a library of $2604$ extragalactic source SEDs to which we apply a pseudo-random value of $R_{\rm V}$ from a normal Gaussian distribution with a mean of $R_{\rm V} = 3.1$ and a standard deviation of $0.2$ in order to simulate random Milky Way lines of sight dust properties (at fixed $p_{B5}D$ value).}
\label{deltam_allps_Rvrange}
\end{figure}

In the present paper, we make use of the parametrisation of the extinction curve from Fitzpatrick (1999) and adopt a $R_{\rm V} = 3.1$. The UV region of the extinction curve ($\lambda < 2700\AA$) is parametrised by a $6$-parameter function that relies on a Lorentzian bump model of the position and width of the $2175\AA$ bump and a far-UV slope term. Along with the Milky Way curve parameters derived for $R_{\rm V} = 3.1$, F99 also provides similar coefficients for $R_{\rm V}$-dependent extinction laws. We probe the impact on Galactic absorption estimates of the adopted value of  $R_{\rm V}$. Figure~\ref{deltam_allps_Rvrange} shows the distribution of Galactic extinction for $p_{B5}D = 0.1$ for the same library of SEDs generated in section 3.5 and for the bluer of our considered bands (for visibility). We respectively adopt $R_{\rm V} = 2.9, 3.1$ and $3.3$. The uncertainty on $R_{\rm V}$ leads to uncertainties $\Delta A_{\rm X}$ on Galactic extinction estimates of the order of $0.04$ mag in the bluer filters. We caution the reader that even the uncertainties on $A_{\rm X}$ due to $R_{\rm V}$ are SED-dependent. 

In order to mimic the potential dust properties that could be encountered along the line of sight of the sources, we also apply to our $2604$ SEDs a random value of $R_{\rm V}$ drawn from a normal Gaussian distribution with a mean of $R_{\rm V} = 3.1$ and a standard deviation of $0.2$. We repeat the exercise a hundred times and average the distribution of the Galactic extinction for a given filter (see Figure~\ref{deltam_allps_Rvrange} bottom panel). These distributions therefore provide a `best' estimate with uncertainties of Galactic extinction when both the source SED and line-of-sight $R_{\rm V}$ are not known a-priori. Gaussian fits were performed on the distributions. Mean $A_{\rm X}$ and standard deviations $\Delta A_{\rm X}$ are provided in Table~\ref{rangeDm2}. We strongly invite the interested reader to use these values for their estimate of Galactic extinction by linearly rescaling the tabulated values of $A_{\rm X}$ and uncertainties by the reddening value in the direction of their source/field. The uncertainties on Galactic extinction due to the unpredictability of $R_{\rm V}$ in the direction of a source have to be added in quadrature to the errors on the photometric measurements. If a prescription of Galactic extinction such as described in Appendix A is implemented, the errors to be added are smaller since $A_{\rm X}$ and $\Delta A_{\rm X}$ are computed for each SED.

\begin{table}
\caption{Galactic extinction predictions that take into account both the lack of knowledge of the source SED and uncertainties of line-of-sight $R_{\rm V}$}
\label{tablegroup}
\centering
\begin{tabular}{l c c}
\hline
Filter	&	$A_{\rm X}$	&	$\Delta A_{\rm X}$ \\
	&	Gaussian fit	&	Gaussian fit		\\
	& 	($1$)		&	($2$)			\\
\hline 
LSST u	&	0.486	&	0.026	\\
LSST g 	&	0.365	&	0.025	\\
PS1 g	&	0.359	&	0.025	\\
DES g	&	0.362	&	0.025	\\
LSST r	&	0.256	&	0.023	\\
PS1 r 	&	0.255	&	0.023	\\
DES r	&	0.242	&	0.022	\\
VIS 		&	0.197	&	0.020	\\
LSST i	&	0.189	&	0.016	\\
PS1 i	&	0.188	&	0.016	\\
DES i 	&	0.177	&	0.015	\\
LSST z	&	0.148	&	0.011	\\
PS1 z	&	0.148	&	0.011	\\
DES z	&	0.135	&	0.009	\\
LSST y	&	0.123	&	0.007	\\
PS1 y	&	0.123	&	0.007	\\
DES Y	&	0.118	&	0.007	\\
Y		&	0.104	&	0.005	\\
J		&	0.079	&	0.004	\\
H		&	0.059	&	0.003	\\
\hline    
\end{tabular}
\begin{list}{}{}
\small{
\item Galactic extinction expressed in mag for line-of-sight $pD = 0.1$.
\item[($1$)] Gaussian fit mean of the distributions in Figure~\ref{deltam_allps_Rvrange} (bottom panel).
\item[($2$)] Gaussian fit standard deviation of the distributions in Figure~\ref{deltam_allps_Rvrange} (bottom panel).
}
\end{list}
\label{rangeDm2}
\end{table}

\begin{figure}
\begin{center}
\includegraphics[width = 11.5cm,bb= 60 30 650 800]{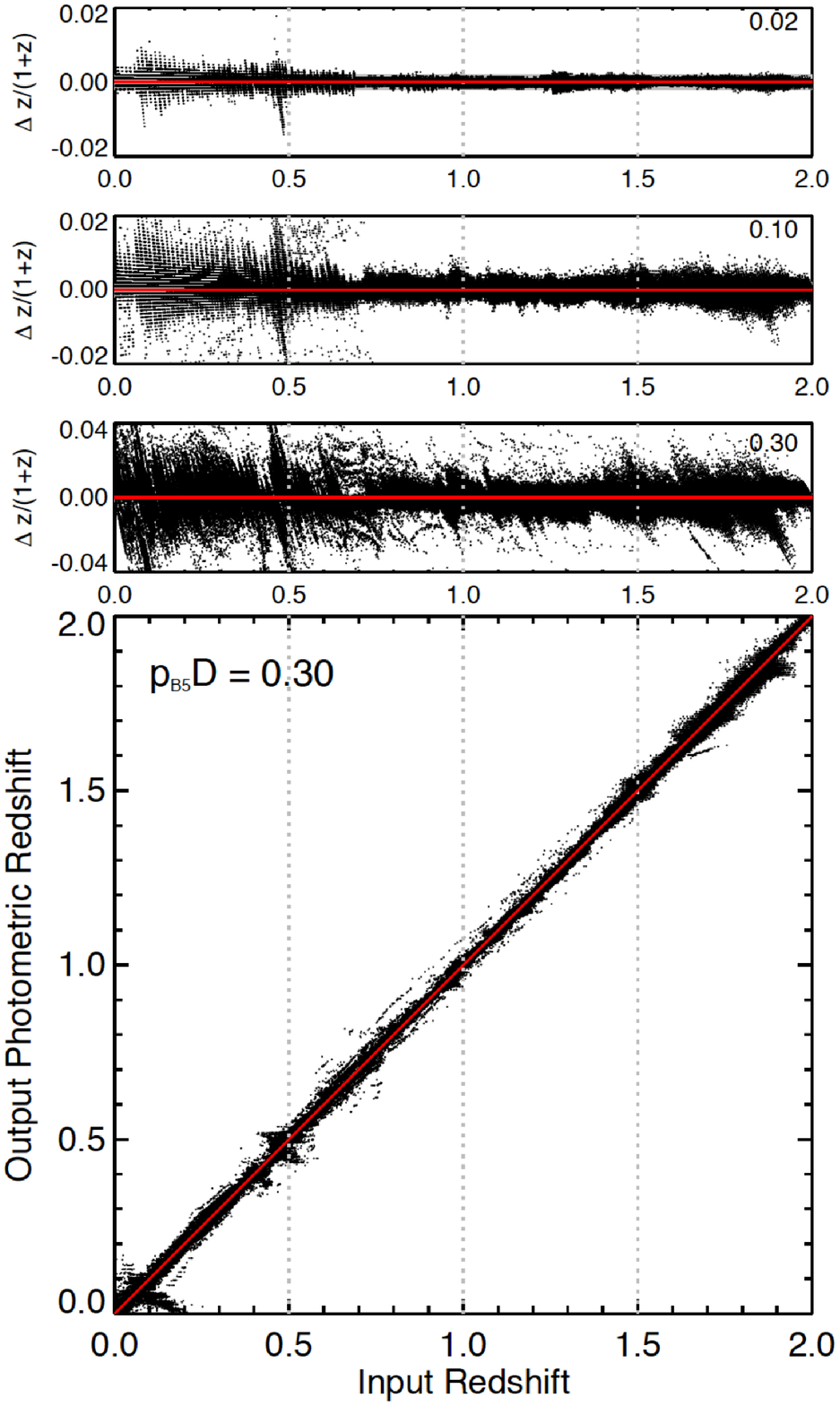}
\end{center}
\caption{Same as Figure~\ref{zp} but this time, the mock catalogue was generated by first reddening the simulated sources by an extinction law whose characteristic $R_{\rm V}$ is allocated randomly from a normal Gaussian distribution with a mean of $3.1$ and a standard deviation of $0.2$. The photometry is then `de-reddened' by the Galactic extinction value derived for a fixed $R_{\rm V} = 3.1$ at fixed $p_{B5}D$. In doing so, we can therefore assess the impact on photo-z of neglecting the uncertainties on $R_{\rm V}$ from the divergences from the identity line.}
\label{zpRv}
\end{figure}

Finally, we evaluate the added errors on photometric redshift estimates induced by the uncertainties on $R_{\rm V}$. Similarly to section 4, we built mock $ugrizyYHJ$ DES $+$ {\it Euclid} catalogues of sources using the same library of templates, intrinsic dust extinction range and redshift grid for three $p_{\rm B5}D$ values ($0.02$, $0.1$ and $0.3$) but this time, we randomly allocate the $R_{\rm V}$ value along the line of sight (from a normal Gaussian distribution with a mean of $3.1$ and a standard deviation of $0.2$) for each simulated source (i.e.~$k_{\rm \lambda}$ changes from source to source). We then correct the reddened mock photometry by the Galactic extinction value derived for a fixed $R_{\rm V} = 3.1$ for the same value of $p_{\rm B5}D$ and within the same filter. In doing so, the observed  discrepancies between photometric redshift and input SED redshift reflect the sole impact of the uncertainties of $R_{\rm V}$ on photo-z. We then attempt to recover the input redshift of the SEDs using Phosphoros with the same SED that were used to generate the catalogues. Figure~\ref{zpRv} shows the differences between the template initial redshifts and the Phosphoros outputs. We note that contrary to the impact of SED on Galactic extinction, the uncertainty on $R_{\rm V}$ do not create strong biases on photo-z estimate. For all investigated values of $pD$, the bias at all redshifts $\langle z \rangle / (1+z)$ is below $0.1\%$ although the average scatter on photo-z estimates of the order of $0.2$\% for $pD = 0.1$ and $0.5$\% for $pD = 0.3$ is non negligible when trying to reach photometric redshifts of high precision.

\section{Recalibrating the $E_{\rm B-V}$ SFD98 map using F99}

As mentioned earlier in the text, the SFD98 $E_{\rm B-V}$ map was derived from the $100\mu$m DIRBE/{\it IRAS}-combined, point-source removed dust map. The dust column density $D$ was linearly rescaled to a reddening value using a scaling factor $p$ which was estimated from the colour excesses of $\sim 500$ local ($z < 0.05$) elliptical galaxies. Contrary to the present work, SFD98 adopted the Milky Way absorption law functional form of \citet{O'Donnell1994} in the visible and Cardelli, Clayton \& Mathis (1989) in the ultraviolet and infrared (CCMOD hereafter). The present appendix presents a recalibration of the DIRBE/{\it IRAS} dust map when adopting a Milky Way absorption law of F99.

\begin{figure*}
\begin{center}
\includegraphics[width = 18.6cm,bb = 5 240 585 465]{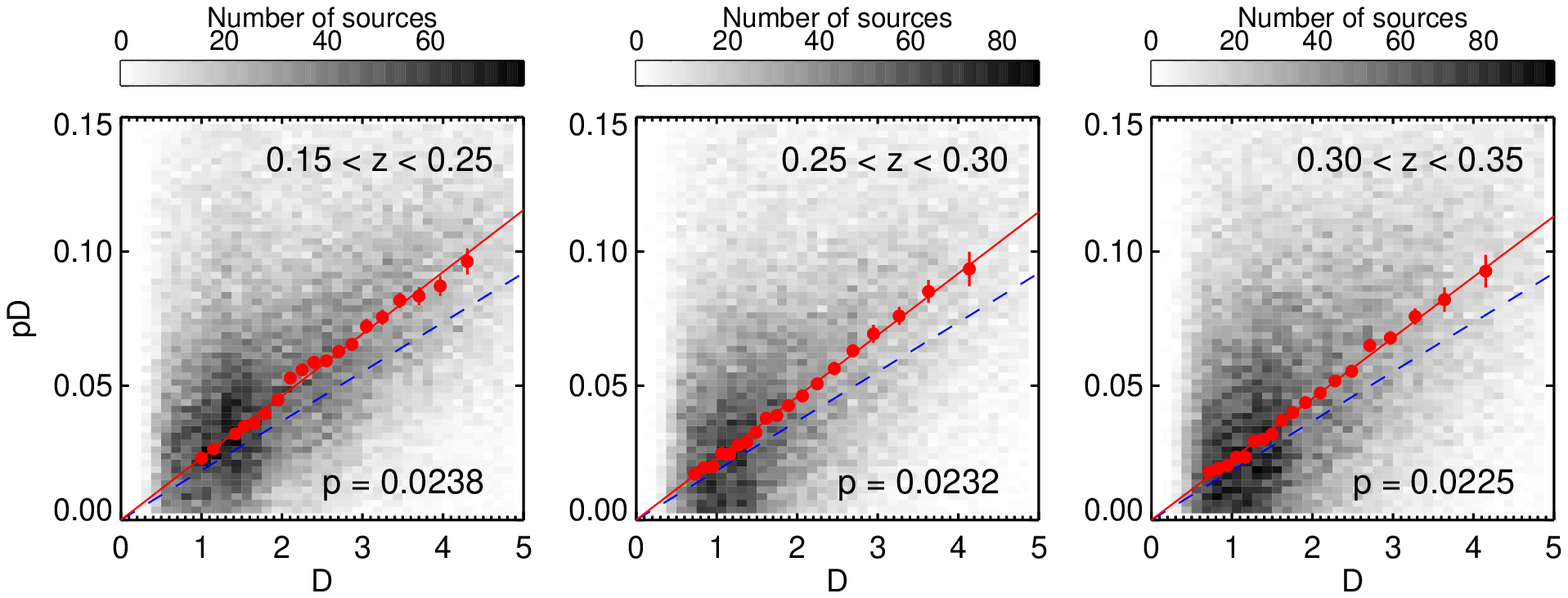}
\end{center}
\caption{Estimates of $pD$ as a function of dust from the DIRBE/{\it IRAS} map measured along the line of sight of SDSS red galaxies at $0.15 < z < 0.25$ (left), $0.25 < z < 0.30$ (middle) and $0.30 < z  < 0.35$ (right). The grey colour maps account for the density of sources. The red dots and error bars represent the median and standard deviation estimated through bootstrapping resampling of $pD$ and $D$ within the range $0.5 < D < 4.5$ split in $20$~bins of equal number of sources. The red line is a linear fit to the dots. We find $p_{\rm LRG} \sim 0.0238$, $0.0232$ and $0.0225$ for $0.15 < z < 0.25$, $0.25 < z < 0.30$ and $0.30 < z < 0.35$ respectively. The blue dashed line shows the $p = 0.0184$ derived by SFD98 using a CCMOD Milky Way absorption law and a local elliptical galaxy sample at ($z < 0.05$).}
\label{boss2}
\end{figure*}

The analysis of section 3.4 where we studied the dependence of the dust-to-reddening factors $p$ with SED using luminous red galaxies was repeated but this time, the values of reddening ($pD$) along the line of sight of our galaxy sample were confronted to the value of the DIRBE/{\it IRAS} dust map $D$. Figure~\ref{boss2} shows the correlation of the $pD$ estimates with $D$. We consider sources in the dust range $0.5 < D < 4.5$. The median $pD$ and $D$ and associated dispersion are derived using bootstrap resampling (see section 3.4). Figure~\ref{boss2} shows the median $(D;pD)$ per bin of dust values in red. We derive the scaling factor $p$ for the different redshift bins using a linear regression and find

\begin{figure}
\begin{center}
\includegraphics[width = 8cm,bb = 30 80 480 480]{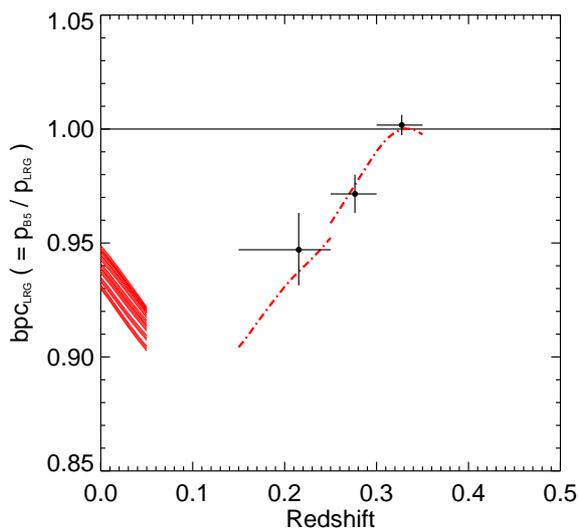}
\end{center}
\caption{Band-pass corrections derived from two red galaxy templates (dotted dashed red lines; see Figure~\ref{converstemp} for details). The $p_{\rm LRG}$ derived from the DIRBE/{\it IRAS} dust map are shown by the black dots with the plotted values corresponding to bpc$_{\rm LRG} = p_{\rm B5} / p_{\rm LRG}$ where the optimal $p_{\rm B5} = 0.0225$. For a direct comparison of this analysis with the Schlegel calibration work, the band-pass corrections derived from local ($z < 0.05$) elliptical galaxy templates from the COSMOS template library (seven templates; see Figure~\ref{conversmodels}) are reproduced here (red solid lines). The average band-pass correction derived from these models is bpc$_{\rm SFD98} = 0.927$.}
\label{converstemp2}
\end{figure}

\begin{itemize} 
\item $p_{\rm LRG[0.15-0.25]} = 0.0238 \pm 0.0004$, 
\item $p_{\rm LRG[0.25-0.30]} = 0.0232 \pm 0.0002$,
\item $p_{\rm LRG[0.30-0.35]} = 0.0225 \pm 0.0001$. 
\end{itemize}

Similarly to Figure~\ref{converstemp}, Figure~\ref{converstemp2} shows the band-pass corrections bpc$_{\rm LRG}$ derived from the two reference templates adopted for our LRG sample to estimate their line-of-sight reddening and the three new values of $p$ derived by the present calibration work based on the DIRBE/{\it IRAS} dust map. The plotted quantities correspond to bpc$_{\rm LRG} = p_{\rm B5} / p_{\rm LRG}$ where the normalisation factor $p_{\rm B5}$ is obtained using a weighted mean (see section 3.4). We find $p_{\rm B5} = 0.0225$. This new analysis confirms once again that the scaling factor estimates and their relative ratios are consistent with what is expected from predictions of band-pass corrections. 

In order to directly confront this new calibration work with the analysis of SFD98, we estimate the scaling factor expected for a sample of local elliptical galaxies. From the COSMOS elliptical galaxy template library (see section 3.2 and Figure~\ref{converstemp2}), we derive that the average band-pass correction bpc$_{\rm SFD98} \sim 0.927$. According to the present calibration work, this would correspond to a dust-to-reddening scaling factor for a local elliptical SED-type of $p_{\rm SFD98} = 0.0243$. SFD98 derived a scaling factor $p = 0.0184 \pm 0.0014$; this discrepancy reflects, among other possible calibration caveats, the impact of using different absorption laws. 

We finally quantify this difference in terms of Galactic extinction in a given filter $X$, $A_X$. We estimate $A_X$ for a set of passive galaxy models from the COSMOS template library for $z = [0,0.05]$ and no intrinsic dust ($E_{\rm B-V}^{Int} = 0$). If a local galaxy is observed through a line of sight $E_{\rm B-V} = 0.1$ (i.e.~the value of the $E_{\rm B-V}$ SFD98 map at that position) in a $g$-band filter (here LSST $g$-band), the Galactic extinction estimate derived at the pivot wavelength of the filter ($0.1 \, k_{\rm pivot,g}$), assuming a CCMOD absorption law, would be $A_g = 0.485$~mag. Taking into account the SED shape of the whole passive SED library, we would derive a range of $A_g = [0.355-0.367]$~mag. At that same position on our newly-calibrated DIRBE/{\it IRAS}-based reddening map calibrated using F99, we would read a reddening $E_{\rm B-V} = 0.1321$ (i.e.~$0.1 \times 0.0243 / 0.0184$) and derive a range of $A_g = [0.462-0.478]$~mag. For an (LSST) $r$-band filter, we would find a CCMOD pivot $A_r = 0.269$~mag, a CCMOD range $A_r = [0.263-0.270]$~mag and a F99 range $A_r = [0.327-0.339]$~mag. We note that this level of uncertainties on Galactic extinction estimates due to the adopted absorption law itself is at the level of the scatter in Galactic extinction estimates expected from the uncertainties of the F99 characteristic total-to-selective ratio parameter $R_{\rm V}$ in Appendix B.

\end{appendix}

\end{document}